\RequirePackage{ifpdf}
\documentclass[11pt,a4paper]{article}
\usepackage{amsmath,amssymb,epsfig,jheppub}

\font\cmss=cmss12

\newcommand\half{\frac12}
\newcommand\del{\partial}
\newcommand\bi{\begin{itemize}}
\newcommand\ei{\end{itemize}}

\newcommand\tl{{\tilde \ell}}

\newcommand\tc{{\tilde c}}
\newcommand\tk{{\tilde k}}

\newcommand\tilh{{\tilde h}}
\newcommand\tchi{{\tilde \chi}}

\newcommand\hchi{{\hat \chi}}

\newcommand\halpha{{\hat \alpha}}
\newcommand\ha{{\hat a}}

\newcommand\bea{\begin{eqnarray}}
\newcommand\eea{\end{eqnarray}}
\newcommand\be{\begin{equation}}
\newcommand\ee{\end{equation}}
\newcommand\nn{\nonumber}

\newcommand\btau{{\bar \tau}}

\newcommand\cN{{\cal N}}

\newcommand\bchi{{\overline \chi}}

\newcommand\sfrac[2]{{\textstyle\frac{#1}{#2}}}

\newcommand\shalf{{\textstyle\frac12}}

\newcommand\ZZ{\hbox{Z\kern-.4emZ}}
\newcommand\sZZ{\hbox{\sevenfont Z\kern-.4emZ}}

\newcommand{\eref}[1]{Eq.\,(\ref{#1})}
\newcommand{\Comment}[1]{{}}

\def\IB{\relax{\rm I\kern-.18em B}}
\def\IC{{\relax\hbox{\kern.3em{\cmss I}$\kern-.4em{\rm C}$}}}
\def\ID{\relax{\rm I\kern-.18em D}}
\def\IE{\relax{\rm I\kern-.18em E}}
\def\IF{\relax{\rm I\kern-.18em F}}
\def\II{\relax{\rm I\kern-.18em I}}
\def\IZ{\mathbb{Z}}
\def\Id{\relax{1\kern-.32em 1}}
\def\IG{\relax\hbox{$\inbar\kern-.3em{\rm G}$}}
\def\IR{\relax{\rm I\kern-.18em R}}

\title{Towards a Classification of Two-Character Rational Conformal Field Theories} \author{A. Ramesh Chandra\footnote{Email:
    ramesh.ammanamanchi@gmail.com} and Sunil Mukhi\,\footnote{Email:
    sunil.mukhi@gmail.com}\\ \it Indian Institute of Science Education and Research,\\ \it Homi Bhabha Rd, Pashan, Pune 411 008, India} 

\abstract{We provide a simple and complete construction of infinite families of consistent, modular-covariant pairs of characters satisfying the basic requirements to describe two-character RCFT. These correspond to solutions of generic second-order modular linear differential equations. To find these solutions, we first construct ``quasi-characters'' from the Kaneko-Zagier equation and subsequent works by Kaneko and collaborators, together with coset dual generalisations that we provide in this paper. We relate our construction to the Hecke images recently discussed by Harvey and Wu.}

\preprint{}


\arxivnumber{1810.09472}

\begin{document}

\maketitle

\section{Introduction}

\label{intro}

The traditional classification of Rational Conformal Field Theories (RCFT) in 2d is based on either (chiral) symmetry algebras or lattices. In the former approach, after selecting a particular chiral algebra or set of algebras, one uses the structure of null vectors to write down a minimal series of CFT's that furnish realisations (not necessarily unitary) of the given algebra \cite{Belavin:1984vu, Knizhnik:1984nr} (for more examples see \cite{DiFrancesco:1997nk} and references therein). Among possible chiral algebras, Kac-Moody algebras are special because one can consider affine models containing all integrable primaries at a given level, namely WZW models, and then take cosets of them. This coset construction generates a vast supply of RCFT's including the various minimal series already obtained by null-vector methods. The approach based on lattices includes the famous $c=1$ classification \cite{Ginsparg:1987eb, Dijkgraaf:1987jta} and the constructions based on even, self-dual lattices as reviewed for example in \cite{Goddard:1983at}.

These two approaches, the chiral algebra/WZW coset approach and the lattice approach, are not complete -- we know RCFT's that belong in neither of these classes \cite{Tuite:2008pt, Hampapura:2016mmz}. Moreover, there are simple coset models which do not come from affine WZW theories \cite{Gaberdiel:2016zke}. Hence it can be useful to organise RCFT's not by these criteria but by simplicity, defined as the number of independent scaling exponents or primary fields, roughly the same as the number of characters.

The $c=24$ models classified in \cite{Schellekens:1992db} are examples of one-character CFT's where there is an extremely large chiral algebra, including many spin-2 generators. The only primary field of this extended algebra is the identity. Several two-character RCFT's have likewise long been known \cite{Mathur:1988na,Naculich:1988xv}. A complete classification of the simplest models could be more useful than the traditional approach for various purposes, including applications in condensed-matter systems as well as for studies from a mathematical viewpoint. An additional application of simple rational CFT's has emerged in the context of four-dimensional supersymmetric gauge theories \cite{Beem:2013sza,Beem:2017ooy,Buican:2017rya}. 

An approach to the classification of RCFT with a small numbers of characters was proposed long ago in \cite{Mathur:1988na}. It is based on the observation that the characters of a rational CFT are vector-valued modular forms of weight zero and satisfy a Modular Linear Differential Equation (MLDE) (this fact was independently noted in \cite{Eguchi:1988wh}). Thus, to classify them we may start from a general class of MLDE and identify values of the parameters for which they determine ``admissible'' sets of candidate characters -- those whose $q$-expansion has non-negative integer coefficients\footnote{Our use of ``admissible'' may differ from some of the mathematical literature.}. After finding admissible characters there are well-defined procedures, using the modular transformation matrix $S$ and its relation to the fusion rules via the Verlinde formula \cite{Verlinde:1988sn}, to determine whether they truly describe a consistent RCFT, and if so to solve the resulting theories. These procedures were implemented in considerable detail in \cite{Mathur:1988gt,Naculich:1988xv} and some recent work in this direction can be found in \cite{Mukhi:2017ugw}. There is also now a considerable mathematical literature on using MLDE to find vector-valued modular forms of CFT type, of which some relevant works are \cite{Bantay:2005vk,Mason:2007,Bantay:2007zz, Bantay:2010uy,Marks:2011,Gannon:2013jua,Arakawa:2016hkg,Arike:2016ana,Franc:2016} and additional ones will be described in what follows.

Modular linear differential equations are characterised by an integer $n$, the number of characters, which is also the order of the differential equation, as well as an integer $\ell\ge 0,\ell\ne 1$ which specifies the number of zeroes in moduli space of the Wronskian determinant of the solutions (this concept will be explained in more detail below). In \cite{Mathur:1988na} all admissible characters with $n=2, \ell=0$ were classified and form a finite set. All but one have been identified with definite RCFT's, while one has a degenerate vacuum and was initially rejected for this reason. It is now understood to be an Intermediate Vertex Operator Algebra (IVOA) \cite{Kawasetsu:2014} based on the notion \cite{Landsberg:2004} of Intermediate Lie Algebra. Related mathematical work on the 2nd order $\ell=0$ equation may be found in \cite{KZ,KK,Kaneko:On, Mason:2008,Kaneko:2013uga,Mason:2018}.

Subsequently the case $n=2,\ell=2$ was classified in \cite{Naculich:1988xv, Hampapura:2015cea}. Again there is a finite set of admissible characters. A remarkable property of these characters, shown in \cite{Gaberdiel:2016zke}, is that they are in one to one correspondence with the $\ell=0$ theories and satisfy a ``generalised coset'' relationship with them. This in particular proves that the $\ell=2$ characters correspond to consistent RCFT's. Like its counterpart at $\ell=0$, one of them has a degenerate vacuum and in this respect resembles an IVOA. We will refer to characters having a degenerate vacuum state as being of ``IVOA type'' although there is no implication that an intermediate Lie algebra as in \cite{Landsberg:2004,Kawasetsu:2014} is involved. 

The status for larger values of $\ell$ is as follows. It was argued in \cite{Naculich:1988xv} and explicitly verified in \cite{Hampapura:2015cea} that there are no admissible characters with $\ell=3$. In fact \cite{Naculich:1988xv} argued that with two characters, there can be no admissible solutions for any odd $\ell$. For $\ell=4$, \cite{Naculich:1988xv} observed that admissible solutions can exist, and searched for them, but then concluded that there are no new ones other than tensor products. More recently, \cite{Hampapura:2015cea} failed to find any models with $\ell=4$ but they did not make an exhaustive search. However, admissible characters with $\ell=4$ do in fact exist. Most of these correspond to tensor product RCFT's, while precisely three of them turn out to be irreducible in this sense and we will describe them below. Two of them first appeared in \cite{Tener:2016lcn}, whose authors restricted their study to potentially unitary CFT's. Out of the three that we find, the first two agree with \cite{Tener:2016lcn} while the third is in the Lee-Yang class, hence the corresponding CFT -- if any -- would be non-unitary. 

To summarise, for two-character theories the parameter $\ell$ labelling the differential equation has to be even, and for $\ell<6$ the classification leads to a finite set of characters for $\ell=0,2$ that have been identified as RCFT's, while for $\ell=4$ there is also a finite set, most of which are tensor products but a few are new admissible characters. They are yet to be identified as RCFT's. 

The classification for $\ell\ge 6$ is more complicated and  has remained largely unexplored until recently. It is known that taking tensor products increases the value of $\ell$, but of course we would like to discover ``irreducible'' theories that are not tensor products. It has long been known \cite{Mathur:1988gt} that irreducible RCFT's with $\ell$ arbitrarily large do exist. A simple example noted in \cite{Hampapura:2015cea} lies on the orbifold line of a free boson theory with $c=1$ at radius $R=\sqrt{2p}$. This theory has $p+4$ characters and $\ell=3(p-1)$ for arbitrary integer $p\ge 1$. However, this series starts at five characters. The first significant insight into this question for two-character theories, from the point of view of classification, was recently provided by Harvey and Wu in \cite{Harvey:2018rdc}. These authors defined a novel type of Hecke operator that acts on vector-valued modular forms to provide new ones. In this process the number of characters is preserved, and under certain conditions the positivity of coefficients in the $q$-expansion is also preserved, so that the Hecke images are admissible characters. These Hecke images have increasing $\ell$ and it is easy to find admissible characters with arbitrarily large values of $\ell$ (as well as the central charge $c$).

In this paper we re-examine the classification problem for $\ell\ge 6$. We find two distinct methods, both based on a notion of ``quasi-characters'' which we introduce, that generate infinite families of new admissible characters with arbitrarily large values of the parameter $\ell$ and the central charge $c$. We are able to demonstrate that the second of these methods (based on addition of quasi-characters) is complete, in that it generates all possible admissible characters with any $\ell$. This solves a major outstanding problem and constitutes a significant step towards the classification of two-character RCFT. To complete the classification process one would have to identify which admissible characters correspond to genuine RCFT's. Partial progress towards this goal can be found in \cite{Chandra:2018ezv}. 

The organisation of this paper is as follows. In Section \ref{sec.onechar} we start by reviewing one-character (often called ``meromorphic'') CFT's and place them in the context of the modular linear differential equation (MLDE) approach. While the results are not new, some of the details are not widely appreciated and will be useful to provide a context for the ensuing discussion of two-character theories. Thereafter in Section \ref{sec.twochar} we discuss in some detail the status of two-character RCFT with $\ell<6$. In addition to the review material there will be some new results that contradict previous claims. We will also highlight several important subtleties. 

In Section \ref{sec.quasichar} we summarise several ways in which characters solving a 2nd-order MLDE can ``fail''. In this category are pairs of characters where one is modular invariant on its own, pairs that give rise to negative but nonetheless integral fusion rules, characters where one achieves integrality of Fourier coefficients only if the vacuum state is chosen to be degenerate, and pairs with exponents that differ by an integer, which correspond to logarithmic CFT. But the most important category for us, neglected in many previous works, are those where all the ``degeneracies'' are {\em integer but not necessarily positive}. We call these ``quasi-characters'' and will find that they are useful building blocks for admissible characters. We go on to show that a certain parametrisation introduced by Kaneko and collaborators \cite{KZ,KK,Kaneko:On} gives rise to infinite families of such quasi-characters with $\ell=0$. We generalise these considerations to find a fresh series of quasi-characters with $\ell=2$. The two sets turn out to obey the coset/bilinear relation of \cite{Gaberdiel:2016zke} just as the genuine admissible characters do.

Finally in Section \ref{sec.mainresult} we present our main results: that infinite sets of admissible characters with arbitrarily large values of $\ell$ (and $c$) can be constructed using quasi-characters. We describe two ways to do this. One of these ways involves multiplication of a quasi-character with a suitable function of the $j$-invariant, and the other involves taking linear combinations of different quasi-characters. We provide a proof that the latter method, applied to quasi-characters with $\ell=0,2,4$, generates all possible admissible characters for any $\ell$. A key outcome of our work is that for $\ell\ge 6$ one can find infinitely many admissible characters with the same central charge, and it is also easy to construct examples with large $\ell$ as well as $c$. We then compare the results of our constructions to the Hecke image approach of \cite{Harvey:2018rdc}. We conclude with a summary and discussion.

\section{One-character CFT}

\label{sec.onechar}

\subsection{Basic features}

We start our discussion with one-character theories because they carry some important lessons. Such theories have a partition function of the form:
\be
Z(\tau,\btau)=|\chi(\tau)|^2
\ee
where $\chi(\tau)$ is holomorphic. Defining $q=e^{2\pi i\tau}$, the character behaves near $\tau\to i\infty$ as $\chi \sim q^{-\frac{c}{24}}$ where $c$ is the central charge. It has long been known that for this partition function to be invariant under $SL(2,\IZ)$, $c$ must be a multiple of 8. For $c=8$ there is a unique character $\chi_{E_{8,1}}$ that is modular invariant up to a phase. The resulting CFT is the WZW model for the algebra $E_8$ at level $k=1$. For $c=16$ there is again a unique character, $\left(\chi_{E_{8,1}}\right)^2$. This time there are two CFT's, namely the WZW models based on the algebras $E_8\oplus E_8$ and Spin$(32)/\IZ_2$. For $c=24$ there are infinitely many characters of the form:
\be
\chi(\tau)= j(\tau)+{\cal N}
\ee
where $j(\tau)$ is the Klein $j$-invariant, which happens to be equal to $\left(\chi_{E_{8,1}}\right)^3$, and ${\cal N}$ is, at this stage, any real number. Note that modular invariance alone puts no condition on ${\cal N}$.

Thus at $c=24$ there appear to be potentially infinitely many characters, depending on the added constant ${\cal N}$. Now we need to invoke a physical requirement. For the resulting partition function to describe a conformal quantum field theory, the coefficients in the expansion of $\chi(\tau)$ in powers of $q$ must describe degeneracies of states. Hence they must be non-negative integers. This is already true of all the coefficients in the expansion of $j$, but can potentially be violated by our choice of the constant ${\cal N}$. Indeed the $q$-expansion of the above character is:
\be
j(\tau)+{\cal N}=q^{-1}+(744+{\cal N})+196884 q+\cdots
\ee
We see that the second coefficient is integer only if ${\cal N}$ is integer, and is non-negative if ${\cal N}\ge -744$. This still leaves us with an infinite set of admissible characters that might describe a CFT, one for every integer ${\cal N}\ge -744$. The next question is whether all of these describe actual CFT's. 

This is precisely the problem that was addressed in 1991 by Schellekens \cite{Schellekens:1992db}, who argued that there are just 71 consistent single-character CFT's with $c=24$. Some features of his result are worth noting here: (i) there are less than 71 different characters that correspond to consistent CFT's, because in many cases a given character describes more than one CFT. This is analogous to the situation at $c=16$ where there is a unique character but two different CFT's, (ii) the minimal value of ${\cal N}$, namely $-744$, is realised as a rather special CFT. In this case the expansion of the character is missing the term of order $q$ relative to the leading term. This means the CFT, if there is one, does not have any Kac-Moody symmetry (because among all possible chiral symmetry algebras of spin-$n$, only spin-1 generators create a secondary at the first level above the identity state). And indeed, not only is there a CFT at this value but it is a celebrated one: the Monster CFT \cite{Frenkel:1988xz,Borcherds:1}.

Beyond $c=24$, as observed for example in \cite{Mathur:1988na}, the single character of a one-character theory must be a function of the Klein $j$-invariant of the form:
\be
\chi(\tau)=j^{w_\rho}(j-1728)^{w_i} P_{w_\tau}(j)
\label{onechar}
\ee
where $w_\rho\in \{0,\frac13,\frac23\}$, $w_i\in \{0,\half\},w_\tau\in\IZ$ and $P_{w_\tau}(j)$ is a polynomial of degree $w_\tau$ in $j$. Notice that the $c=8$ and $c=16$ characters arise by choosing $\delta=\frac13, \frac23$ respectively with $\beta=0$ and $P_{w_\tau}(j)=1$. Clearly there is an infinite number of potential characters with $c>24$, and by examining the behaviour as $q\to 0$ we find the central charge of the corresponding CFT (if there is one) would be 
\be
c=24(w_\rho+w_i+w_\tau)
\label{cfromnr}
\ee
But as we already noted in the case of $c=24$, only a subset of these will be admissible as characters, namely those whose Fourier coefficients are all non-negative integers (in particular this subset has $w_i=0$). By carefully choosing the polynomial one can find the admissible characters. The next step would be to repeat the procedure of Schellekens and find out how many of these correspond to consistent CFT's. Unfortunately for $c\ge 48$ this seems a very hard problem and remains open at present.

\subsection{Modular linear differential equations}

Let us now phrase the above results in the language of differential equations. Though essentially trivial, this will give a nice introduction to the integer $\ell$ mentioned above, and set the stage for the the study of 2-character theories for arbitrary $\ell$. We start by noting that for any of the 1-character theories, the single character satisfies the trivial equation:
\be
\left(D  - \Big(\frac{D\chi(\tau)}{\chi(\tau)}\Big)\right)\chi(\tau)=0
\label{trivdiff}
\ee
where $D\equiv\frac{1}{2\pi i}\frac{\del}{\del \tau}$. The purpose of writing this equation is that it provides a way to ``experimentally re-discover'' the characters. To do this, we rewrite the above equation as:
\be
\big(D+  \phi_2(\tau)\big)\chi(\tau)=0
\label{firstord}
\ee
Since the character is modular invariant (up to a phase) it has weight 0 under $SL(2,\IZ)$. It is easy to see that differentiating in $\tau$ increases the modular weight by 2. Hence for the equation to be invariant, the function $\phi_2(\tau)$ must transform with weight 2 under $SL(2,\IZ)$ modular transformations:
\be
\phi_2\left(\frac{a\tau+b}{c\tau+d}\right)=(c\tau+d)^2 \phi_2(\tau)
\ee
One may think that $\phi_2$ has to be holomorphic, but in fact it is generically meromorphic. Referring to \eref{trivdiff}, we see that $\phi_2$ has a pole whenever $\chi$ has a zero. And despite $\phi_2$ having such a pole, the solution of the differential equation is perfectly holomorphic (away from $\tau\to i \infty$, as always).

We now classify the differential equation by the number of zeroes of $\chi$. This is generically the same as the number of poles of $\phi_2=-\frac{D\chi}{\chi}$, though sometimes the zeroes of $\chi$ could be partly cancelled by zeroes of $D\chi$. Because of the nature of the torus moduli space where $\tau$ takes values, the number of poles of any meromorphic function is quantised as $\frac{\ell}{6}$ where $\ell=0,2,3,4,\cdots$. The reason why fractional powers arise is that the space has orbifold points at $\tau=\rho\equiv e^{\frac{2\pi i}{3}}$ of order 3, and $\tau=i$ of order 2. As a result, one can have poles of order $\frac13$ and $\half$ respectively at these points, corresponding to $\ell=2,3$ respectively. One gets $\ell=4$ by having coincident poles at $\rho$, and $\ell=5$ by having a pole each at $i$ and $\rho$. Finally with $\ell=6$ one can have a pole anywhere in moduli space (we call this a ``bulk pole''). Bulk poles can be present in principle for all higher values of $\ell$ except $\ell=7$.

It will be useful to note that all modular forms under $SL(2,\IZ)$ are polynomials in the Eisenstein series $E_4$ and $E_6$ of weight 4 and 6 respectively. If one wants meromorphic forms, they can be constructed by taking ratios. Due to the location of the zeroes of $E_4$ and $E_6$, when they appear as factors in the denominator they contribute $\ell=2$ and $\ell=3$ respectively. 

Since the moduli space is compact, the total number of zeroes and poles of the character must be equal. This is a special case of the Riemann-Roch theorem. The pole is only at infinity, where we assume the behaviour $\chi\sim q^\alpha$. We have the identification $\alpha=-\frac{c}{24}$. Meanwhile the number of zeroes is just $\frac{\ell}{6}$.  It follows that:
\be
\alpha=-\frac{\ell}{6}~ \hbox{or equivalently } c=4\ell
\label{lalpha}
\ee
which from \eref{cfromnr} tells us that $\frac{\ell}{6}=w_\rho+w_i+w_\tau$. 

Now let us try to build \eref{firstord} starting with the simplest case, $\ell=0$. There are no holomorphic modular forms of weight 2, so we must have $\phi=0$. The unique solution is $\chi=1$ (we are free to choose the normalisation). This is the trivial character. The next case is $\ell=2$. In this case we must take $\phi_2=\mu\frac{E_6}{E_4}$, where for the moment, $\mu$ is an arbitrary real coefficient. Since the $q$-expansion for the Eisenstein series is known, the differential equation can be solved order by order in $q$ and gives rise to a character that is modular invariant (up to a phase) for every value of the free parameter $\mu$. The resulting character will have a power-series expansion of the form:
\be
\chi(\mu,\tau)=q^{\alpha}(1+a_1(\mu)q+a_2(\mu)q^2+\cdots)
\ee
Note that we have chosen the coefficient of the leading term to be 1. This is because the corresponding state is the vacuum state of the theory, which is expected to be unique. Since $\ell=2$, from \eref{lalpha} we have $\alpha=-\frac13$, so the central charge of the associated CFT (if it exists) is $c=8$. Solving recursively to the first two orders, we find:
\be
\mu=-\alpha,\quad a_1(\mu) = 744\mu 
\ee
which in turn tells us that $\mu=\frac13$ and $a_1=248$. Subsequent orders will determine $a_2(\mu),a_3(\mu)$ etc. The remarkable fact is that all of these numbers are non-negative integers. Moreover, the value of $a_1$ determines the dimension of the Kac-Moody algebra of the theory, in this case 248. Thus we have rediscovered the $E_{8,1}$ character. 

One can repeat this exercise for $\ell=3$ by writing the equation:
\be
\left(D+\mu \frac{E_4^2}{E_6}\right)\chi=0
\ee
This time, Riemann-Roch tells us that $\alpha=-\frac12$ which means $c=12$. Recursively solving the equation gives:
\be
\mu=-\alpha=\half,\quad a_1(\mu)=-984\mu=-492
\ee
In this case we see that at the first level above the ground state, the ``degeneracy'' is a negative integer. This fails the test to be an admissible character and therefore there is no one-character CFT with $c=12$. Nevertheless, it turns out that this case satisfies a weaker consistency condition: the degeneracies, although sometimes negative, are nonetheless all integers. In what follows we will find that such ``quasi-characters'' can be useful as building blocks, though by themselves they do not describe any CFT.

It is straightforward to continue in this way. At $\ell=4$ one finds a unique character having $c=16$ and $a_1=496$, which is the dimension of $E_8\oplus E_8$ as well as Spin$(32)/Z_2$. The case of $\ell=5$ has $\phi_2=\mu_1 \frac{E_6}{E_4}+\mu_2\frac{E_4^2}{E_6}$. Again we find that the character has integral coefficients, but some are negative. At $\ell=6$ we have $\phi_2=\mu_1\frac{E_4^2 E_6}{E_4^3+\mu_2 \Delta}$ where $\Delta=\frac{E_4^3-E_6^2}{1728}$ is the modular discriminant (in our conventions). As expected, this has a ``bulk pole'' where $E_4^3+\mu_2\Delta=0$. This case has $\alpha=-1$, corresponding to $c=24$. Inserting this and solving the equation recursively, we find:
\be
\mu_1=-\alpha=1,\quad a_1(\mu_i)=744+\mu_2
\ee
Continuing to solve for the character to one higher order, we find that $a_2=196884$, independent of the value of $\mu_2$, and all subsequent terms are non-negative integers and independent of $\mu_2$. From the above equation it is clear that to have an admissible character we must take $\mu_2=N$, an integer $\ge -744$. Indeed, the character in this case is simply $\chi_{\ell=6}=j(\tau)+N$. As stated above, there are 71 consistent CFT's having characters in this set, while all the remaining (infinitely many) characters of this form seem to be unrelated to any CFT. Beyond $\ell=6$, one can generate admissible characters using the polynomial \eref{onechar}, setting $w_i=0$, $w_\rho\in 0,\frac13,\frac23$ and choosing the coefficients suitably in the polynomial $P_{w_\tau}(j)$. In this way one can generate admissible characters for any arbitrarily large, even $\ell=6(w_\rho+w_\tau)$.

The reader may have noticed that we could have simply integrated the differential equation \eref{firstord} instead of solving it by a series expansion in $q$. Indeed there are far simpler ways to extract the above results on admissible characters, and \eref{onechar} summarises the complete set of results for the one-character case, which have long been known in the mathematical literature -- so there was no need to resort to differential equations in the first place. But the reason we have gone through this exercise is to highlight a number of features  that will for the most part generalise to the two-character case, where a general solution analogous to \eref{onechar} does not exist. The relevant features of the one-character case are:
\bi
\item
there are admissible, holomorphic characters for arbitrarily large values of $\ell$, 
\item
when $\ell$ lies in the range $0\le \ell<6$, there is a finite set of admissible characters for $\ell=0,2,4$ and none for $\ell=1,3,5$, 
\item
every admissible character in this range corresponds to a consistent CFT, \item
for $\ell\ge 6$ one encounters situations where the characters depend on arbitrary integers and are admissible over infinite ranges of these integers. However, most of them do not correspond to genuine CFT's. 
\ei

\section{Two characters: complete classification for $\ell<6$}

\label{sec.twochar}

\subsection{The general theory and the map to hypergeometric equations}
\label{ell.less.6}

Analogous to \eref{firstord} for the one-character case, we can write the general two-character differential equation. A new feature relative to the one-character case is that we need to use covariant derivatives, defined by:
\be
D_{(k)}\equiv \frac{1}{2\pi i}\del - \frac{k}{12}E_2(\tau)
\ee
where $k$ is a positive even integer and $E_2(\tau)$ is the second Eisenstein series, which is a holomorphic connection on moduli space. It transforms inhomogeneously under modular transformations, such that if $f_{(k)}(\tau)$ transforms with weight $k$ under $SL(2,\IZ)$ then:
\be
 D^n_{(k)}\equiv D_{(k+2n-2)} D_{(k+2n-4)}\cdots D_{(k+2)} D_{(k)} f_{(k)}(\tau)
\ee
transforms with weight $k+2n$. We denote $D_{(0)}$ simply by $D$. The special case $D^n_{(0)}$ is the only one relevant for CFT characters, since those have zero weight\footnote{Note however that mathematical works such as \cite{KZ} do make use of $D_k^{(n)}$ for specific choices of $k$.}.
Hence it will be simply written $D^n$.
With these definitions, the general MLDE for two characters is:
\be
\Big(D^2 +\phi_2(\tau)D + \phi_4(\tau)\Big)\chi=0
\label{twochar}
\ee
where $\phi_2,\phi_4$ are meromorphic and modular of weight 2,4.
The characters that solve the equation take the form:
\be
\begin{split}
\chi_0 &= q^{\alpha_0}\left(1+m_1q+m_2q^2+\cdots\right)\\
\chi_1 &= \mathrm{D}\, q^{\alpha_1}\left(1+a_1 q+a_2 q^2+\cdots\right)
\end{split}
\label{twocharsoln}
\ee
Notice that we have treated the two characters somewhat asymmetrically. The first one, which we associate to the identity character, has the exponent $\alpha_0=-\frac{c}{24}$. It is normalised so that its series expansion starts with 1, since this is the vacuum degeneracy and one normally expects a non-degenerate vacuum state. The series coefficients $m_i$ then have to be non-negative integers in order to be interpreted as degeneracies. The second character corresponds to the non-trivial primary field, and has the exponent $\alpha_1=-\frac{c}{24}+h$ where $h$ is the (chiral) conformal dimension of the primary\footnote{The $\alpha_i$'s emerge as the zeroes of a quadratic equation, so a choice has to be made about which root is associated to the identity character. The preceding equations show that for unitary theories where $h>0$, the lower root should be identified with $\alpha_0$. We will see later that this identification needs to be re-examined if the theory is non-unitary.}. The second character is normalised by multiplying with a positive integer D which is the degeneracy of the primary state. The expansion coefficients $a_i$ do not need to be integers. It is sufficient if they are rational numbers with a denominator that divides D, so that the products D$\,a_i$ are integers.

If one knows the characters then in principle one can compute the modular transformation matrix $S$ defined as follows:
\be
\chi_i\left(-\frac{1}{\tau}\right)=\sum_{j=0}^1 S_{ij}\chi_j(\tau)
\label{smatrixdef}
\ee
If this matrix turns out to be unitary then the partition function:
\be
Z(\tau,\btau)= \bchi_0(\btau)\chi_0(\tau)+\bchi_1(\btau)\chi_1(\tau)
\ee
will be modular invariant. However, it is also possible that the non-identity character has a multiplicity M, corresponding to the fact that a single character corresponds to more than one primary. Such will be the case when there is some discrete symmetry such as complex conjugation that causes the primaries to appear in pairs or larger multiples. In this case, the modular-invariant partition function will be:
\be
Z(\tau,\btau)= \bchi_0(\btau)\chi_0(\tau)+\mathrm{M}\,\bchi_1(\btau)\chi_1(\tau)
\ee
It is important to distinguish the degeneracy D from the multiplicity M. The former tells us only that a particular primary is degenerate, the latter says that there are M different primaries, each with degeneracy D, whose character is the same. In the presence of a multiplicity, the matrix $S_{ij}$ will not be unitary but rather will satisfy:
\be
S^\dagger \begin{pmatrix} 1&0\\ 0&M\end{pmatrix} S=\begin{pmatrix} 1&0\\ 0&M\end{pmatrix}
\label{Smult}
\ee

If the characters $\chi_0,\chi_1$ were known, the coefficient functions $\phi_2,\phi_4$ in the differential equation would be determined in terms of the Wronskians:
\be
W_0=\left|\begin{matrix} \del\chi_0 & \del\chi_1\\
\del^2\chi_0 & \del^2\chi_1\end{matrix}\right|,\quad
W_1=\left|\begin{matrix} \chi_0 & \chi_1\\
\del^2\chi_0 & \del^2\chi_1\end{matrix}\right|,\quad
W=\left|\begin{matrix} \chi_0 & \chi_1\\
\del\chi_0 & \del\chi_1\end{matrix}\right|
\ee
as:
\be
\phi_2=-\frac{W_1}{W}, \quad \phi_4=\frac{W_0}{W}
\ee

The differential equation is labelled by the number $\frac{\ell}{6}$ of zeroes of the Wronskian $W$ (in the one-character case of Section \ref{sec.onechar} the Wronskian was just the character $\chi$) which in turn labels the maximum allowed poles of the modular forms $\phi_2,\phi_4$. We will examine the solutions to this requirement below. But before we turn to that question, let us observe some general features. Suppose we consider the subfamily of differential equations where the poles are located only at $\rho,i$ (as explained above, this is the general case for $\ell<6$ while it is a special case for $\ell\ge 6$). Then, as studied in \cite{Naculich:1988xv, Mathur:1988gt}, the differential equation \eref{twochar} can be mapped to a hypergeometric equation and its solutions explicitly found in terms of hypergeometric functions. \cite{Naculich:1988xv} went further and used the Klein $j$-invariant itself, known as a ``hauptmodul'' of $SL(2,\IZ)$, as the variable for the differential equation. In terms of this, one can actually bypass the differential equation and write the characters as functions of $j$. Suppose the characters behave near the points $\tau=\rho,i,i\infty$ where $j\to 0,j\to 1728,j\to\infty$ as:
\be
\begin{split}
\chi_0(\tau)&\sim j^{r_\rho},\quad \sim (j-1728)^{r_i},\quad ~j^{-\alpha_0}\\
\chi_1(\tau)&\sim j^{s_\rho},\quad \sim (s-1728)^{s_i},\quad ~j^{-\alpha_1}
\end{split}
\ee
where $r_\rho,s_\rho\ge 0$ and $\in \IZ/3$, while $r_i,s_i\ge 0$ and $\in \IZ/2$. Then, the Riemann-Roch theorem tells us that:
\be
\alpha_0+\alpha_1=\frac{1-\ell}{6}
\ee
Equating $\alpha_0,\alpha_1$ to $-\frac{c}{24}, -\frac{c}{24}+h$ respectively, where $h$ is the chiral conformal dimension associated to the character $\chi_1$, the above relation becomes:
\be
-\frac{c}{12}+h=\frac{1-\ell}{6}
\ee

Defining:
\be
\begin{split}
w_\rho=r_\rho+s_\rho-\sfrac13,\qquad & w_i=r_i+s_i-\sfrac12,\\
t_\rho=s_\rho-r_\rho,\qquad &t_i=s_i-r_i
\end{split}
\label{rsdefs}
\ee
it was shown in \cite{Naculich:1988xv} that:
\be
\begin{split}
&\frac{\ell}{6}=w_\rho+w_i,\\
&t_\rho,t_i>0,~t_\rho,t_i \notin \IZ 
\label{wtdefs}
\end{split}
\ee
and the characters can be expressed in terms of $j$ as: 
\be
\begin{split}
\chi_0(j)&=j^{\frac{c}{24}}\left(1-\frac{1728}{j}\right)^{r_i}F\left(\alpha,\beta,\gamma;\frac{1728}{j}\right)\\
\chi_1(j)&=\sqrt{m}j^{\frac{c}{24}-h}\left(1-\frac{1728}{j}\right)^{r_i}F\left(\alpha+1-\gamma,\beta+1-\gamma,2-\gamma;\frac{1728}{j}\right)
\end{split}
\label{genhyper}
\ee
where:
\be
\alpha=-\shalf(h+t_\rho+t_i-1),\quad 
\beta=-\shalf(h-t_\rho+t_i-1),\quad 
\gamma=1-h
\ee
and
\be
m=(1728)^{2h}\frac{\sin\pi\alpha\,\sin\pi\beta}{\sin\pi(\alpha-\gamma)\,\sin\pi(\gamma-\beta)}
\left(\frac{\Gamma(\gamma)\Gamma(1-\alpha)\Gamma(1-\beta)}{\Gamma(2-\gamma)\Gamma(\gamma-\alpha)\Gamma(\gamma-\beta)}\right)^2
\ee
In terms of the degeneracy D and multiplicity M introduced above, one sees that $m=\mathrm{MD}^2$. Armed with these results, we first summarise all cases with $\ell<6$.

\subsection{The case of $\ell=1,3,5$}

The first equation in \eref{wtdefs} already tells us that $\ell=1$ is not possible, given that the RHS is either 0 or else has a minimum value of $\sfrac13$. Requiring that there are two independent solutions when expanded about the singular points $\rho, i$, imposes $t_\rho,t_i$ $\notin \IZ$ (if this were not true, then we would find a logarithmic behaviour at either of the special points). This immediately implies, using \eref{rsdefs}, that $w_i \in \IZ$ and $\ell \in 2\IZ$. Thus, when $\ell$ is an odd integer, the space of solutions to the MLDE is one dimensional. This rules out two-character CFT with odd $\ell$. Note that this argument doesn't change when there are additional zeroes elsewhere in the moduli space, the only condition being imposed is that the critical exponents at the singular points do not differ by an integer.

\subsection{The case of $\ell=0$}

The first attempt at classifying MLDE was carried out in \cite{Mathur:1988na} for the case of two characters and $\ell=0$. The map to the hypergeometric equation was not used there and was in fact proposed later, independently in \cite{Mathur:1988gt} and \cite{Naculich:1988xv}. From the above discussion it is clear that for $\ell=0$, the forms appearing in \eref{twochar} are $\phi_2=0$ and $\phi_4=\mu E_4$ where $\mu$ is a free parameter. Hence the equation is:
\be
\Big(D^2  + \mu E_4(\tau)\Big)\chi=0
\label{ellzero}
\ee
We will refer to this as the MMS equation\footnote{Note a change of normalisation in the definition of the parameter $\mu$ relative to that in \cite{Mathur:1988na}, namely $\mu_{\rm MMS}=-4\mu_{\rm here}$.}. As shown there, the parameter $\mu$ in the equation is related to the central charge as $\mu=-\frac{c(c+4)}{576}$.

Using the map to hypergeometric equations described in subsection \ref{ell.less.6}, we find that the general solutions are:
\be
\begin{split}
\chi_0(\tau) &= j^{\frac{c}{24}}{}_2F_{1}\left(-\frac{c}{24},\frac{1}{3}-\frac{c}{24};\frac{5}{6}-\frac{c}{12} ;\frac{1728}{j}\right) \\
\chi_1(\tau) &= \sqrt{m}\,j^{-\frac16-\frac{c}{24}}{}_2F_{1}\left( \frac{1}{6}+\frac{c}{24},\frac{1}{2}+\frac{c}{24};\frac{7}{6}+\frac{c}{12};\frac{1728}{j}\right)
\end{split}
\label{ellzero.hyper}
\ee
This can then be used to generate the $q$-expansion. We now summarise the results. For generic rational values of $\mu$, the $q$-expansion of the solutions of \eref{ellzero} has rational ``degeneracies'' but the rational numbers have growing denominators as we go up in the order of expansion. Such solutions are to be eliminated. However, for some special values of $\mu$, the expansion coefficients become integer or rational with a bounded denominator. Some of these cases involve negative integer coefficients and were also eliminated in previous works\footnote{These are the ``quasi-characters'' that will play an important role later in the present work.}. At this stage, precisely 10 solutions survive -- each having the property that (possibly after choosing an overall integral normalisation) every coefficient in the $q$-expansion, up to a large order, is a non-negative integer. The 10 solutions in \cite{Mathur:1988na} are listed in Table \ref{table-leq.0}.

This is admittedly an empirical method, wherein integrality is verified to a high power of $q$ and then conjectured to hold to all orders. Its success lies in the fact that at least in the $\ell=0$ case, each of the resulting characters can be given a definite physical interpretation in the language of CFT. This interpretation then amounts to a ``physics proof'' of the desired integrality to all orders. There are also other ways to prove integrality without having to identify with a concrete CFT, these will be discussed below.

%\begin{table}[ht]
\begin{table}
\centering
{\renewcommand{\arraystretch}{1.4}
\begin{tabular}{|c||c|c|c|c|c|c|c|}
\hline
No. & $\mu$ & $c$ & $h$ & $m_1$ & D & M & Theory\\
\hline
1 & $\frac{11}{3600}$ & $\frac{2}{5}$ & $\frac15$ & 1 & 1 &1 & Lee-Yang \\
\hline
2 & $\frac{5}{576}$ & $1$ & $\frac14$ & 3 & 2 &1 & $ A_{1,1}$\\
\hline
3 & $\frac{1}{48}$ & $2$ & $\frac13$ & 8 & 3 &2 & $A_{2,1}$\\
\hline
4 & $\frac{119}{3600}$ & $\frac{14}{5}$ & $\frac25$ & 14 & 7 &1 & $G_{2,1}$\\ 
\hline
5 & $\frac{2}{36}$ & 4 & $\frac12$ & 28 & 8 & 3 & $D_{4,1}$ \\
\hline
6 & $\frac{299}{3600}$ & $\frac{26}{5}$ & $\frac35$ &52 & 26 &1 & $F_{4,1}$\\
\hline
7 & $\frac{5}{48}$ & $6$ & $\frac23$ & 78 & 27 & 2 & $E_{6,1}$\\
\hline
8 & $\frac{77}{576}$ & 7 & $\frac34$ & 133 & 56 & 1 & $E_{7,1}$\\
\hline
9 & $\frac{551}{3600}$ & $\frac{38}{5}$ & $\frac45$ & 190 & 57 &1 & $E_{7.5,1}$ \\
\hline
10 & $\frac{1}{6}$ & 8 & $\frac56$ & 248 & $-$ & 0 & $E_{8,1}$\\
\hline
\end{tabular}}
\caption{The MMS Series.}
\label{table-leq.0}
\end{table}
%\end{table}

Of these 10 pairs of characters, the most straightforward are the seven cases numbered 2-8. These are Wess-Zumino-Witten (WZW) models at level $k=1$ for the corresponding Lie algebra. It is remarkable that the MLDE approach, at these specific values of $\mu$, gives us complete information about the characters of these models, which can be independently obtained from the famous Kac-Weyl formula (see for example \cite{Kac}). These straightforward cases will help us highlight a few important features. In every case, the number $m_1$, defined in \eref{twocharsoln}, corresponds to the dimension of a simple Lie algebra. Here, $\chi_0$ is the identity character. Recall that each character has its leading exponent, thus for two character theories one has $\alpha_0,\alpha_1$. As mentioned earlier, the natural choice is to take $\alpha_0$ to be the smaller root of the equation. In these cases this choice is reaffirmed by the consistent identification of the characters with known, unitary RCFT's. We see that for the non-identity character, the leading term corresponds to a positive integer degeneracy D (see \eref{twocharsoln}) which, in every case, turns out to be the dimension of the fundamental representation of the associated Lie algebra. In the approach of \cite{Mathur:1988na} this fact was not assumed, but rather it emerged upon seeking the minimum degeneracy factor that ensures integrality of all the coefficients in the $q$-expansion of the non-identity character.

Notice that in some cases the multiplicity M defined in Section \ref{ell.less.6} is different from 1. For $A_2$ and $E_6$ we have M=2. This arises from the fact that the fundamental representation is complex. Also for $D_4$ we have M=3, associated to the triality between the vector, spinor and conjugate spinor representations, which ensures that all have the same character. These multiplicities are not required to make the $q$-expansion integral, so one might ask how they were discovered in the MMS approach. In fact this was done in subsequent works, \cite{Mathur:1988gt,Naculich:1988xv}, where the associated characters were computed exactly in terms of hypergeometric functions. Using these, the modular transformation matrix $S$ defined in \eref{smatrixdef} was computed. Four of them turned out to be unitary while the remaining three did not. Instead they satisfied \eref{Smult} with M=2 in two of the cases, and M=3 in the final case. 

Finally, the fusion rules for these theories are computed using the Verlinde formula \cite{Verlinde:1988sn}:
\be
N_{ijk}=\sum_{m=0}^1 \frac{S_{im}S_{jm}S_{km}}{S_{0m}}
\label{Verlinde}
\ee
It is a rather miraculous feature of the Verlinde formula that the left-hand-side turns out to be a non-negative integer in all consistent CFT's, while the right-hand side is made from a matrix whose entries are not even rational in general. This was verified for each of the above seven theories in \cite{Mathur:1988gt}.

We must now consider the remaining three cases, numbers 1, 9 and 10 in the table. Each of these teaches us a new lesson and leads to valuable insights for the general case. Let us first address entry No.10. Notice that in the table we have chosen the multiplicity of the non-trivial primary to be 0. What does this mean? It turns out that the WZW model for $E_8$ at level 1 has a single character. This is related to the fact that in $E_8$, the fundamental and adjoint representations are the same. In fact we already encountered this theory in our discussion of one-character theories in a previous Section. It is obvious that any modular-invariant set of $p$ characters will not only solve a differential equation of degree $p$, but also differential equations of every degree $p'>p$. In the latter case, we will find $p$ characters that transform among themselves under $SL(2,\IZ)$ as well as $p'-p$ ``spurious'' characters. What we are seeing in entry No.10 is such a recurrence of the $E_8$ character, as part of a 2-character set where the second character is spurious. This means that under modular transformations, the first character is invariant (up to a phase) on its own and does not mix with the second.

Assembling the entries 2-8 and 10 in the table, we find the level-1 WZW models for a series of Lie algebras $A_1, A_2, G_2, D_4, F_4,E_6, E_7,E_8$. The MLDE approach unites them despite their having different Kac-Moody algebras. Subsequently Deligne \cite{Deligne:1} proposed that the corresponding Lie algebras form a series with special representation-theoretic properties (similar observations have been made by Cvitanovi\' c \cite{Cvitanovic:2008zz}). 

Let us now turn to the remaining cases, entries Nos.\,1 and 9 of the table. If the first one were a CFT it would seem to contradict some familiar wisdom: all unitary theories with $c<1$ are minimal models \cite{Belavin:1984vu}, but there is no unitary minimal model with $c=\frac25$. So the theory, if it exists, must be non-unitary. This by itself is acceptable since we are not seeking to classify only unitary theories. However, computation in \cite{Mathur:1988gt} of the fusion rules using \eref{Verlinde} revealed a more serious problem: some of the fusion rule coefficients were found to be {\em negative integers}. Concretely, the modular transformation matrix for this theory is:
\be
S=\begin{pmatrix}\sqrt{\frac{2}{5-\sqrt5}} &
\sqrt{\frac{2}{5+\sqrt5}} \\
\sqrt{\frac{2}{5+\sqrt5}} &
-\sqrt{\frac{2}{5-\sqrt5}} \end{pmatrix}
\label{unitS}
\ee 
Inserting this in \eref{Verlinde} leads to:
\be
N_{000}=N_{011}=1,\quad N_{111}=-1
\ee
This is inconsistent as it stands, since the fusion rules of a consistent RCFT (even non-unitary) are supposed to be non-negative integers.

The remedy, as noted in \cite{Mathur:1988gt}, is to exchange the roles of the two characters. The exponents $\alpha_i$ for this theory are $-\frac{1}{60}$ and $\frac{11}{60}$. While solving the MLDE, it was assumed that the former is to be identified with $-\frac{c}{24}$, leading to the values in the table. Suppose however that we identify the latter exponent with $-\frac{c}{24}$. In this case, we find that $c=-\frac{22}{5}, h=-\frac{1}{5}$. These are the correct values for the famous (non-unitary) Lee-Yang minimal model! The matrix $S$ now has its rows and columns permuted and becomes: 
\be
S=\begin{pmatrix}-\sqrt{\frac{2}{5-\sqrt5}} &
\sqrt{\frac{2}{5+\sqrt5}} \\
\sqrt{\frac{2}{5+\sqrt5}} &
\sqrt{\frac{2}{5-\sqrt5}} \end{pmatrix}
\label{nonunitS}
\ee
The result is merely a change of sign of some of the entries, but it has the desired effect. The fusion rule coefficients are now:
\be
N_{000}=N_{011} =N_{111}=1
\ee
We will refer to the original ordering of characters ($\alpha_0<\alpha_1$) as the ``unitary presentation'', and the reverse ordering ($\alpha_1<\alpha_0$) as the ``non-unitary presentation''. Note that a given theory will be consistent only in one of these two presentations. If it is consistent in the unitary presentation, it still needs to be verified whether it is unitary or non-unitary (the only thing that is assured in the unitary presentation is that the conformal dimension $h$ is positive, this is a necessary but not sufficient condition for the theory to be unitary). However if a theory is consistent in the non-unitary presentation, it is necessarily non-unitary -- since in this case $h$ is negative. (Meanwhile, somewhat confusingly, the matrix $S$ is unitary in both presentations.)

To summarise what we have learned about entry No. 1 of the table, this is inconsistent in the unitary presentation but potentially consistent (the fusion rules are non-negative) in the non-unitary presentation. Moreover in the latter presentation the corresponding RCFT is known to exist and is a well-known minimal model. Note that one could consider exchanging characters in the other cases (Nos.\,2-8), then one would find a system with some negative fusion rules. Our consistent policy, which will continue in the following sections, is to classify only theories with a positive central charge -- these may of course be either unitary or non-unitary. Then, only if the fusion rules turn out to be negative, we consider the non-unitary presentation and check if the fusion rules have now become positive.

We now move on to entry No.\,9 of the table. The matrix $S$ of this theory is precisely the same as that of entry No.\,1, thus in the unitary presentation it is given by \eref{unitS} while in the non-unitary presentation it is the one in \eref{nonunitS}. It follows that the fusion rules are inconsistent in the unitary presentation and consistent in the non-unitary presentation. This tempts us to simply exchange the characters, as we did in case No.\,1. Unfortunately we encounter a new problem having to do with the degeneracy factor D. In entry No.\,1, the degeneracy of the non-identity character (in the unitary presentation) turned out to be 1. This was fortuitous, and ensured that upon exchanging characters to go to the non-unitary presentation the vacuum of the theory continues to be non-degenerate. Unfortunately in entry No.\,9, one finds D=57. Thus in the non-unitary presentation the identity character, and hence the vacuum of the theory, is degenerate. For this reason, entry No.\,9 is inconsistent as a CFT both ways: in one presentation the fusion rules are negative while in the other, the vacuum is degenerate. Hence it was rejected in \cite{Mathur:1988gt}. Nonetheless it is interesting that this case appears to lie ``between'' the $E_7$ and $E_8$ WZW models -- but there is no WZW model occupying such a position and even within finite-dimensional Lie algebras, no mathematical structure of this sort was known at the time. 

However the situation changed in 2004 when it was proposed \cite{Landsberg:2004} to augment the Deligne series by the inclusion of a structure called $E_{7\half}$ that lies ``between'' $E_7$ and $E_8$. This was argued to fill a ``hole'' in Deligne's classification. The numerology of $E_{7\half}$ matches beautifully with entry No.\,9 in the MMS series, for example the number of states above the identity character (in the unitary presentation) is 190, corresponding to the ``dimension'' of $E_{7\half}$, suggesting that what was found as entry No.\,9 in \cite{Mathur:1988na} is a kind of generalised CFT associated to $E_{7\half}$. More recently it has been proposed \cite{Kawasetsu:2014} that these characters be considered as those of an ``Intermediate Vertex Operator Algebra'' (IVOA) based on $E_{7\half}$\footnote{The same work also showed that entry No.\,1 of Table \ref{table-leq.0} can be viewed an IVOA corresponding to the algebra ``$A_\half$'' which is intermediate between the trivial algebra and $A_1$.}. An IVOA has some, but not all, of the features of a Vertex Operator Algebra (VOA) which is the mathematically precise formulation of conformal field theory. So far it is not understood whether IVOA can be thought of as a generalised kind of RCFT in a physically useful way. If so, one would still need to decide whether it should be considered in the unitary presentation (giving up positivity of the fusion rules, but retaining a non-degenerate vacuum) or in the non-unitary presentation (retaining positive fusion rules but allowing a degenerate vacuum). 

There is one more exotic case, not in the table, that is worth mentioning. In  \cite{Mathur:1988na} it was noticed that the first-level degeneracy $m_1$ of the identity character diverges at $c=10$. In recent years it has been understood  that such theories have logarithmic characters\cite{Flohr:1995ea,Flohr:1996vc}, where one of the characters (which we take to be the identity character) has a power-series expansion of the form:
\be
\chi_0(q)=\chi_0^{(1)}(q)\log q+\chi_0^{(2)}(q)
\ee
In the context of the mathematical literature on MLDE this was noted in \cite{Kaneko:2013uga}, building on previous results of \cite{KK,Kaneko:On}. It was shown in \cite{Kaneko:2013uga} that for such logarithmic solutions to the $\ell=0$ equation, the coefficient functions have non-negative integer coefficients in their $q$-expansion for the values $c=10,22,34,46,58,106,226$. The authors described this kind of system as a ``non-rational Vertex Operator Algebra''. 

Logarithmic characters arise whenever the conformal dimension of the primary field becomes an integer. At $\ell=0$, it is easily seen from the Riemann-Roch equation that this happens whenever $c=2(6m+5)=10,22,34,\cdots$ which includes all the cases found in \cite{Kaneko:2013uga}. A particularly simple example arises at $c=-2$, where the two characters are:
\be
\chi_0=\frac{1}{2\pi i}\log q\, \eta^2(q),\qquad \chi_1=\eta^2(q)
\ee
and it is easily verified that they are exchanged by modular transformations. Logarithmic CFT's are interesting in their own right, but also because they bound certain ``regions'' in parameter space where regular CFT's are found as we will see below. 

The reason we have spent considerable space discussing exotic types of characters: non-unitary, IVOA type and logarithmic, is that they occur repeatedly in investigations of MLDE. The first two will play a particularly important role in our discussion of quasi-characters below. 

\subsection{The case of $\ell=2$}

This case was considered in \cite{Naculich:1988xv, Kiritsis:1988kq, Hampapura:2015cea, Gaberdiel:2016zke}. The relevant differential equation is now:
\be
\left(D^2+\frac13\frac{E_6}{E_4}D+\mu E_4\right)\chi=0
\label{leq.2}
\ee
The coefficient $\frac13$ in front of the second term was determined from the indicial equation about $\tau=i\infty$, and the parameter $\mu$ is related to the central charge via $\mu=-\frac{c(c-4)}{576}$. The method of solution is as before, either via $q$-series or via hypergeometric functions. We write down the solution in the latter form:
\be
\begin{split}
\chi_0(\tau) &= j^{\frac{c}{24}}{}_2F_{1}\left(-\frac{c}{24},\frac{2}{3}-\frac{c}{24};\frac{7}{6}-\frac{c}{12} ;\frac{1728}{j}\right) \\
\chi_1(\tau) &= \sqrt{m}\,j^{\frac16-\frac{c}{24}}{}_2F_{1}\left( -\frac{1}{6}+\frac{c}{24},\frac{1}{2}+\frac{c}{24};\frac{5}{6}+\frac{c}{12};\frac{1728}{j}\right)
\end{split}
\label{elltwo.hyper}
\ee
One now proceeds by expanding this in a $q$-series. The resulting admissible characters are shown in Table \ref{table-leq.2}. 

\begin{table}[ht]
%\TABLE{
\centering
{\renewcommand{\arraystretch}{1.4}
\begin{tabular}{|c||c|c|c|c|c|c|}
\hline
No. & $\mu$ & $c$ & $h$ & $m_1$ & D & M \\
\hline
1 & $-\frac13$  & $16$ & $\frac76$ & 496 & $-$ & 0 \\
\hline
2 & $-\frac{1271}{3600}$ & $\frac{82}{5}$ & $\frac65$ & 410& 902 &  1 \\
\hline
3 & $-\frac{221}{576}$ & 17 & $\frac54$ & 323 & 1632 & 1\\ 
\hline
4 & $-\frac{7}{16}$ & 18 & $\frac43$ & 234 & 2187 & 2   \\
\hline
5 & $-\frac{1739}{3600}$ & $\frac{94}{5}$ & $\frac75$ & 188 & 4794 & 1\\
\hline
6 & $-\frac{5}{9}$ & 20 & $\frac32$ &140 & 5120 &  3\\
\hline
7 & $-\frac{2279}{3600}$ & $\frac{106}{5}$ & $\frac85$ & 106 & 15847 &  1\\
\hline
8 & $-\frac{11}{16}$ & 22 & $\frac53$ & 88 & 16038 &  2\\
\hline
9 & $-\frac{437}{576}$ & 23 & $\frac74$ & 69 & 32384 & 1 \\
\hline
10 & $-\frac{2891}{3600}$ & $\frac{118}{5}$ & $\frac95$ & 59 & 32509 & 1\\
\hline
\end{tabular}}
\caption{Consistent CFT's with $\ell=2$.}
\label{table-leq.2}
%}
\end{table}

We have not included a ``Theory'' column in this table, not because the information is unavailable but because it is too long to summarise in the available space. As shown in \cite{Gaberdiel:2016zke} in the context of the novel coset construction described there, the seven theories numbered 3-9 in the table correspond to very special non-diagonal invariants for certain direct-sum Kac-Moody algebras. All of them are cosets of meromorphic theories of \cite{Schellekens:1992db} by one of the seven ``standard'' WZW models in the original MMS series, namely those appearing in entries 2-8 of Table \ref{table-leq.0}. A little inspection reveals that entries 3-9 of Table \ref{table-leq.2} pair up (in reverse order) with entries 2-8 of Table \ref{table-leq.0} in such a way that 
\be
c+\tc=24,\qquad h+\tilh=2
\label{cosetsum}
\ee
where $(c,h), (\tc,\tilh)$ are the central charge and conformal dimension in Tables\,\ref{table-leq.0} and \ref{table-leq.2} respectively. This property is a basic feature of the coset construction of \cite{Gaberdiel:2016zke}, where it is also shown that coset duals of theories with $\ell=0$ have $\ell=2$ and vice-versa. The coset property is defined in terms of the Sugawara construction in \cite{Gaberdiel:2016zke} but can also be expressed through the bilinear relation:
\be
\chi_0(\tau)\tchi_0(\tau)+\chi_1(\tau)\tchi_1(\tau)=j(\tau)-744+\cN
\label{bilin}
\ee
where $\cN$ is a non-negative integer that varies from case to case.

As happened with $\ell=0$, again we seem to have found seven clearly consistent sets of characters, which all correspond to RCFT's via the coset construction of \cite{Gaberdiel:2016zke}. Note in particular that the fusion rules of the cosets are identical to those of the original theories.

What about the remaining three entries of Table \ref{table-leq.2}? The pattern is very much the same as in Table \ref{table-leq.0}. Again, we have three ``outliers'' which we now describe in turn. Let us start with No.\,10. In the unitary presentation, this has central charge $\tc=24-\frac25$ and primary dimension $\tilh=\frac95$. If we compare it with entry No.\,1 of Table \ref{table-leq.0}, we see that  $c+\tc=24,h+\tilh=2$.  This again suggests the coset construction of \cite{Gaberdiel:2016zke}, and indeed case No.\,10 of Table \ref{table-leq.2} is the (generalised) coset dual of case No.\,1 of Table \ref{table-leq.0}. The reason we call this ``generalised'' is that these are not theories with conventional Kac-Moody algebras. Instead, the coset relation is defined via the bilinear relation \eref{bilin}, which can be verified even in this case, between the characters of the two theories. An important point to note here is that this coset structure holds when both theories are considered in the unitary presentation. However, in this presentation entry No.\,1 of Table \ref{table-leq.0} has negative fusion rules. Hence the same is true of the dual, entry No.\,10 of \ref{table-leq.2}. 

If instead we switch to the non-unitary presentation on both sides, the first theory becomes consistent, as seen above (with non-negative fusion rules) and equal to a minimal model. However its dual No.\,10 of Table \ref{table-leq.2} does not. This is because, as seen from the table, the non-identity character has a degeneracy of 32509. Thus after interchange of characters this would be a set of characters of Intermediate Vertex Operator Algebra type. Notice that in the non-unitary presentation the two theories would satisfy a different condition from the standard coset condition \eref{cosetsum}, namely $c+\tc=-24, h+\tilh=-2$. For now we are only dealing with admissible characters (non-negative integer coefficients in the $q$-series), but if one steps outside this class then it is possible to find duals for non-unitary presentations \cite{Hampapura:2016mmz}.

Next, consider entry No.\,2 of Table \ref{table-leq.2}. This entry, with $c=\frac{82}{5}$, clearly pairs up with entry No.\,9 of Table \ref{table-leq.0} in the same way as in the above cases, so it can again be considered the (generalised) coset dual of the IVOA associated to $E_{7\half}$, with the coset dual relationship being understood in the sense of \eref{bilin}. Here both theories are of ``IVOA type'', since their degeneracy factors are 57 and 902 respectively. 

Entry No.\,1 of Table \ref{table-leq.2} has an identity character that is modular invariant (up to a phase) by itself, so the second character is spurious. Indeed this is just the character of the famous $E_8\times E_8$ and Spin$(32)/Z_2$ WZW models. This is clearly the coset dual of entry No.\,10 of Table \ref{table-leq.0}, this follows from the fact that $\chi_{E_8}^3=j(\tau)$. Here both theories are completely consistent unitary CFT's and they exhibit a trivial version of the coset relationship.

Finally we note that logarithmic characters exist for $\ell=2$ as well. The Riemann-Roch theorem tells us this happens at specific integer values of the central charge $c=2(6n+1)=2,14,26,\cdots$. These could be deserving of further study, however that will be outside the scope of the present work.

\subsection{The case of $\ell=4$}
\label{ell.4}

Here we shall complete the classification of all two character theories with $l=4$ started in \cite{Naculich:1988xv,Kiritsis:1988kq,Hampapura:2015cea}. In this case the MLDE is:
\be
\left(D^2+\frac23\frac{E_6}{E_4}D+\mu E_4-384\frac{\Delta}{E_4^2}\right)\chi=0
\ee
Again the coefficient of the second term is determined by the indicial equation (in fact it is always equal to $\frac{\ell}{6}$). The coefficient of the last term is determined by considering the indicial equation around $\tau=\rho$. That leaves one free parameter, $\mu$, related to the central charge by $\mu=-\frac{c(c-12)}{576}$.

We definitely expect to find some admissible characters, namely products of the form $\chi_{E_8}\chi_i$ where $\chi_i$ have $\ell=0$. These products all have $\ell=4$ as one can see from the formula for the $\ell$-value of the tensor product of two theories with $(p,p')$ characters and $\ell$-values $(\ell,\ell')$ respectively \cite{Hampapura:2015cea}:
\be
{\tilde \ell}=\half pp'(p-1)(p'-1)+\ell'p+\ell p'
\ee
The question is whether there are any additional admissible characters that are not of tensor-product form.

The value $l=4$ corresponds to the Wronskian having a double zero at the orbifold point $\rho = e^{\frac{2\pi i}{3}}$, and no zeros elsewhere in the moduli space. The characters written as functions of $j(\tau)$ take the following form (up to normalisation):
\be
\begin{split}
\chi_0(\tau) &= j^{\frac{c}{24}}{}_2F_{1}\left( \frac{1}{3}-\frac{c}{24},\frac{2}{3}-\frac{c}{24};\frac{3}{2}-\frac{c}{12} ;\frac{1728}{j}\right) \\
\chi_1(\tau) &= j^{\half-\frac{c}{24}}{}_2F_{1}\left( -\frac{1}{6}+\frac{c}{24},\frac{1}{6}+\frac{c}{24};\frac{1}{2}+\frac{c}{12};\frac{1728}{j}\right)
\end{split}
\label{ellfour.hyper}
\ee

We have chosen $c$ as the free parameter above, and from the Riemann-Roch formula we know that $h = \frac{c}{12} -\half$. Now, performing a $q$-expansion of the above characters, we can derive the following expression 
\be
m_1 = \frac{-5c^2+ 306c-4608}{c-18}
\ee
which gives:
\be
(5c)^2+ 5c(m_1-306)= 90(m_1-256)
\ee
Thus, we see that $5c$ is an integer, such that the discriminant of the above quadratic equation is a perfect square. This restricts $c$ to 34 values. 
Next we calculate $m_i$ for $i \geq 2$ in terms of $c$. For $i=2$ we have:
\be
m_2 = \frac{25 c^4-3105 c^3+117324 c^2-1355292 c+6635520}{2 (c-30) (c-18)}
\ee
Requiring that $m_i$ for $i \geq 2$ be a non-negative integer further cuts down the list. This leaves us with $15$ possibilities, shown in Table \ref{table-leq.4}.

\begin{table}[ht]
%\TABLE{
\centering
{\renewcommand{\arraystretch}{1.4}
\begin{tabular}[t]{|c||c|c |c|c|c|c|}
\hline
No.&$c$ & $h$ & $m_1$&D&M& Remarks \\ \hline
1&6& 0 & 246& 1 &  &  logarithmic\\ \hline
2&8 & $\frac{1}{6}$ & 248& 1 &0 &  one-character $E_{8,1}$ \\ \hline
3&$\frac{42}{5}$ & $\frac{1}{5}$ & 249& 1&1 &  $E_{8,1}\otimes YL$\\ \hline
4&9 & $\frac{1}{4}$ & 251&2 &1 & $E_{8,1}\otimes A_{1,1}$ \\ \hline
5&10&$\frac{1}{3}$ & 256 &3 &2 & $E_{8,1}\otimes A_{2,1}$\\ \hline
6&$\frac{54}{5}$ & $\frac{2}{5}$ & 262 & 7& 1& $E_{8,1}\otimes G_{2,1}$\\ \hline
7&12& $\frac{1}{2}$ & 276 &8 & 3& $E_{8,1}\otimes D_{4,1}$\\ \hline
8&$\frac{66}{5}$& $\frac{3}{5}$ & 300&26 & 1& $E_{8,1}\otimes F_{4,1}$ \\ \hline
9&14& $\frac{2}{3}$ & 326 &27 & 2& $E_{8,1}\otimes E_{6,1}$\\ \hline
10&15& $\frac{3}{4}$ & 344 & 56& 1& $E_{8,1}\otimes E_{7,1}$\\ \hline
11&$\frac{78}{5}$& $\frac{4}{5}$ & 438 &57 & 1& $E_{8,1}\otimes E_{7.5,1}$\\ \hline
12&16& $\frac{5}{6}$ & 496  &1 & 0&  one-character $E_{8,1}\otimes E_{8,1}$\\ \hline
13&$\frac{162}{5}$& $\frac{11}{5}$ & 4 & 310124&1 &  A\\ \hline 
14&33& $\frac{9}{4}$ & 3&565760 & 1 &  A \\ \hline
15&34& $\frac{7}{3}$ & 1 & 767637 & 2 &  A\\ \hline
\end{tabular}}
\caption{Potentially consistent CFT's with $\ell=4$.}
\label{table-leq.4}
%}
\end{table}

As indicated, entries No.\,3-12 in the table are the expected tensor products. Hence we may focus our attention on the remaining five cases. Entry No.\,2 is a re-appearance of the one-character theory for $E_{8,1}$. Entry No.\,1 has $h=0$ and is a logarithmic CFT, falling in the series $c=12m+6$ which corresponds to logarithmic characters at $\ell=4$.

This leaves, as the interesting cases, entries No.\,13-15 which have been labelled ``A''. All three were missed in \cite{Naculich:1988xv}, while No.\,13 and 14 (but not 15) appeared implicitly in \cite{Kiritsis:1988kq}, and No.\,14 and 15 (but not 13) have appeared in the mathematics literature \cite{Tener:2016lcn}. No.\,13 with $c=\frac{162}{5}$ is in the same fusion-rule category as the Lee-Yang theory and hence, in the given unitary presentation, has negative fusion rules. Upon reversing the roles of the characters, one finds a non-unitary theory with a 310124-fold degenerate identity character, thus in our language it is a candidate theory of IVOA type. We claim that Nos.\,13-15 form the complete set of admissible $\ell=4$ characters that are not tensor products of other admissible characters.  

\subsection{Fusion rules}

As discussed above, the modular transformation matrix defined in \eref{smatrixdef} and the fusion rule coefficients are related by Verlinde's formula \eref{Verlinde}. In all the above cases, the characters have been determined in terms of hypergeometric functions and this enables us to compute $S_{ij}$ explicitly, from which $N_{ijk}$ may be obtained. A complete classification of possible fusion-rule classes for theories with low numbers of characters was found in \cite{Christe:1988xy,Mathur:1989pk}.  Here we compile the results for all known two-character theories with $\ell<6$. For the $\ell=0$ series, some of the fusion rules were found in \cite{Verlinde:1988sn} and more complete results were presented in \cite{Mathur:1988gt}. Remarkably, as noted in \cite{Gaberdiel:2016zke} and revisited in \cite{Harvey:2018rdc}, the pattern of fusion rules for $\ell=2$ theories follows that of the $\ell=0$ case. We find altogether five different cases that apply to the theories in Tables \ref{table-leq.0} and \ref{table-leq.2}:

(i) For entries No.\,1,4,6,9 of Table \ref{table-leq.0}, entries No.\,2,5,7,10 of Table \ref{table-leq.2} (recall that the latter are cosets paired with the former) and entry No.\,17 of Table \ref{table-leq.4}, one finds that there is one primary $\phi$ (other than the identity) satisfying the fusion rule:
\be
N_{000}=N_{110}=N_{111}=1
\label{fusioncat.1}
\ee
(we follow the convention that any coefficient that is not explicitly written vanishes). These fusion rules are said to be of ``Lee-Yang type''. They correspond to case ${\cal A}_1^{(2)}$ of \cite{Christe:1988xy}.

(ii) For entries No.\,2,8 of Table \ref{table-leq.0}, their cosets No.\,3,9 of Table \ref{table-leq.2}, and No.\,18 of Table \ref{table-leq.4}, we have one primary $\phi$ satisfying the rule:
\be
N_{000}=N_{110}=1
\label{fusioncat.2}
\ee
These fusion rules are said to be of $A_{1,1}$ type. They correspond to case ${\cal A}_1^{(1)}$ of \cite{Christe:1988xy}.

(iii) For entries No.\,3,7 of Table \ref{table-leq.0}, their cosets No.\,4,8 of Table \ref{table-leq.2}, and No.\,19 of Table \ref{table-leq.4},
there are two non-trivial primaries which we label 1 and 2. They are complex conjugates of each other and have the same character. In this case, as analysed in detail in \cite{Mathur:1988gt}, the $2\times 2$ $S$-matrix obtained by considering the modular transformations of the two independent characters is not unitary. Hence, to get the fusion rules, one has to enlarge it manually to a $3\times 3$ matrix, which does turn out to be unitary, and which can be used in \eref{Verlinde} to extract the fusion rules. We find:
\be
N_{000}=N_{120}=N_{112}=N_{221}=1
\label{fusioncat.3}
\ee
These fusion rules are of ${\cal B}_2^{(1)}$ type in the notation of \cite{Christe:1988xy}.

(iv) For entry No.\,5 of Table \ref{table-leq.0} and its coset dual, entry No.\,6 of Table \ref{table-leq.2}, there are three non-trivial primaries. In the $\ell=0$ case these are straightforwardly identified with the vector, spinor and conjugate spinor representations of $D_{4,1}$. However in the coset dual $\ell=2$ case this identification is not so straightforward, since the theory is not a WZW model but rather a non-diagonal invariant of either $(D_{4,1})^5$ or $(A_{5,1})^4$. The fusion rules are:
\be
N_{000}=N_{110}=N_{220}=N_{330}=N_{123}=1
\ee
These fusion rules are of type ${\cal A}_3^{(1)}$ in the classification of  \cite{Christe:1988xy}. We observe that there is no new $\ell=4$ theory with these fusion rules.

(v) For entry No.\,10 of Table \ref{table-leq.0}, entry No.\,1 of Table \ref{table-leq.2} and entries No.\,2,12 of Table \ref{table-leq.4} there are no non-trivial primaries. In the first case this is the $E_{8,1}$ theory and in the second, the $E_{8,1}\times E_{8,1}$ or Spin$(32)/Z_2$ theory. Both cases re-appear at $\ell=4$. The fusion rules are trivially:
\be
N_{000}=1
\ee
These can be called $E_{8,1}$ type fusion rules.

In what follows, instead of the notation of \cite{Christe:1988xy}, we will label the fusion-rule classes (i)-(iv) above as Lee-Yang, $A_1$, $A_2$ and $D_4$ respectively. Notice that we have realised all the fusion rules in \cite{Christe:1988xy,Mathur:1989pk} that can possibly correspond to two-character theories. Any pairs of characters with $\ell\ge 6$ that one may find, will also fall into one of these fusion classes.

\subsection{Comparison to previous work}

As promised in the Introduction, we would like to summarise the aspects in which this section disagrees with stated or implied results in the literature. First, \cite{Kiritsis:1988kq} claimed to classify all two-character theories and argued that there are no irreducible admissible characters for $\ell>0$. This claim is incorrect\footnote{Nevertheless, \cite{Kiritsis:1988kq} contained some interesting observations regarding the methodology of constructing admissible characters that are similar to the present work, and in some ways anticipated our ``product construction'' of subsection \ref{subsec.products}.} and already disagreed with previous work \cite{Naculich:1988xv} which, however, only constructed admissible characters for $\ell=2$. The corresponding RCFT's were constructed in \cite{Gaberdiel:2016zke}, decisively proving that two-character theories with $\ell>0$ do exist. In the present Section we have listed all two-character theories in Table \ref{table-leq.2}.

Continuing to $\ell=4$, \cite{Naculich:1988xv} attempted a classification of these but found only the ``reducible'' ones that are direct products of $\ell=0$ characters with the $E_8$ character, missing the three irreducible cases listed as entries 13-15 in our Table \ref{table-leq.4}. Meanwhile \cite{Hampapura:2015cea} examined the $\ell=4$ MLDE but did not use all available inputs to fix the coefficients. Upon choosing a parameter arbitrarily in the MLDE, no irreducible admissible characters were found -- perhaps conveying the incorrect impression that there are no new admissible characters with $\ell=4$.
Finally \cite{Tener:2016lcn} examined the $\ell=4$ case in particular, and found entries 14 and 15, but did not mention entry 13. It may be noted that to date, all the cases 13-15 only remain admissible characters and there is no known construction of a corresponding RCFT for them.

\section{Quasi-characters}

\label{sec.quasichar}

\subsection{How characters fail}

From the examples above, we see that the classification of characters via MLDE occasionally throws up pairs of characters that are admissible in the limited sense of having non-negative integer coefficients, possibly after choosing a suitable overall normalisation, but still fail one or other criterion. We have seen multiple ways in which they can fail, which we summarise here:

\begin{itemize}
\item
At least one of the fusion rule coefficients is negative. However, after reversing the characters one finds a non-unitary theory that is completely consistent.
\item
At least one of the fusion rule coefficients is negative. After reversing the characters one finds a non-unitary theory whose fusion rules are non-negative, but the identity character is degenerate. We referred to such theories as being ``of IVOA type''.
\item
One of the characters is modular invariant by itself (possibly up to a phase). This is not a failure as such, but rather a re-discovery of a known character together with a ``spurious'' character that has to be discarded.
\item
The conformal dimension $h$ of the non-identity character equals an integer. This is a logarithmic theory.
\end{itemize}

It turns out that the most interesting type of ``failure'' is, however, different from any of the above. There are sets of characters that we rejected from the outset because one or more of the coefficients in the $q$-series, despite being integer, is {\em negative}. While admissible characters are finite in number for fixed $\ell$ (as displayed in Tables \ref{table-leq.0}, \ref{table-leq.2}, \ref{table-leq.4}), there are non-admissible characters whose coefficients are all integers but some are negative, and remarkably there turn out to be {\em infinite numbers of such characters} (a finite subset of these were obtained in \cite{Kiritsis:1988kq}). We will shortly find a use for them in classifying admissible characters with $\ell\ge 6$. Accordingly, we propose the following:

{\bf Definition:} ``Quasi-characters'' are a set of solutions to an MLDE having all integer (but not necessarily positive) coefficients in the $q$-expansion, after selecting a suitable normalisation. Clearly quasi-characters include admissible characters as a special case. But, in the context of second-order MLDE, they are far more numerous. For example, while the original work \cite{Mathur:1988na} found that there are at most 10 admissible solutions to the $\ell=0$ equation, there are infinitely many more solutions that correspond to quasi-characters. As we will now see, the complete set for $\ell=0$ can be extracted from \cite{KK,Kaneko:On} and we will also provide a different complete set for $\ell=2$. 

Within the classification of quasi-characters, we find two distinct types. For one class, which we denote Type I, a finite number of coefficients in the identity character are negative after normalising the ground state to be positive. One encounters all of the negative coefficients below a certain level, above which all the coefficients are positive integers. Meanwhile the other character has all positive coefficients. A different class, which we denote Type II, has a finite number of positive coefficients in the identity character when the ground state is again normalised to be positive. Thus, if we parametrise the identity character as:
\be
\chi_0=q^{-\frac{c}{24}}(a_0+a_1q+\cdots a_nq^n+\cdots)
\label{idchar}
\ee
with $a_0>0$, then for the Type I class we have that some of $a_1,a_2,\cdots a_n$ are $\le 0$, while all of the $a_m > 0$ for $m>n$. For the Type II class we find instead that some of $a_1,a_2,\cdots a_n\ge 0$ while all of the $a_m<0$ for $m>n$. Again, the second character is entirely non-negative. We did not find any examples with a random or alternating collection of plus and minus signs all the way to infinity (which would have fallen outside our classification above). We also find that negative signs occur only in the vacuum character (the character having the most singular behaviour), while the other character has a completely non-negative $q$-series. 
In fact, the asymptotic positivity or negativity of the $q$-coefficients is determined by the modular transformation properties of quasi-characters, more precisely by the modular $S$ matrix. Using the relevant results in Section \ref{modular.quasi} and the asymptotic form of the Rademacher series for these coefficients (as explained for example in \cite{Dijkgraaf:2000}), we give a proof in Appendix \ref{Rademacher} of the empirical facts stated above\footnote{In a previous version of this manuscript, this result was stated as a conjecture. We thank an anonymous referee for urging us to provide a proof, which turns out to perfectly confirm the conjecture.}.

Let us pause to note that in the context of first-order MLDE, it is easy to generate Type I quasi-characters, for example $j(q)+N$ where $N$ is any negative integer $<-744$. One also finds Type II quasi-characters, the simplest one being $(j(q)-1728)^\half$, whose expansion is:
\be
\begin{split}
(j(q)-1728)^\half &= q^{-\half} (1 - 492 q - 22590 q^2 - 367400 q^3 - 3764865 q^4\\
&\qquad \qquad - 28951452 q^5 - 
 182474434 q^6 - 990473160 q^7\\
 &\qquad\qquad - 4780921725 q^8 -\cdots)
\end{split}
\ee
For second-order MLDE, objects equivalent to our quasi-characters have been discussed in mathematical works by Kaneko and collaborators, although not precisely in these words nor in the context to which we will be applying them. In the following subsections, we review these mathematical works, which all deal with the $\ell=0$ second-order MLDE, and generalise them to the $\ell=2$ MLDE. In the following Section we describe how to use quasi-characters to construct admissible characters with arbitrary $\ell\ge 6$.

\subsection{Kaneko-Zagier parametrisation}

\subsubsection*{$\ell=0$ case}

In 1998, Kaneko and Zagier \cite{KZ} studied a variant of the MMS differential equation in the context of supersingular elliptic curves and their invariants. Their equation was for weight-$k$ (rather than weight-0) modular forms, and took the form:
\be
\left(D^2_{(k)}-\frac{k(k+2)}{144}E_4(\tau)\right)f_{(k)}(\tau)=0
\label{KZorig}
\ee
This is easily transformed to MMS form by the substitution:
\be
f_{(k)}(\tau)=\eta(\tau)^{2k}\chi(\tau)
\label{scaledsoln}
\ee
This results in:
\be
\left(D^2-\frac{k(k+2)}{144}E_4(\tau)\right)\chi(\tau)=0
\label{KZeqn}
\ee
which, as we see, is the MMS equation \eref{ellzero} with 
\be
\mu=-\frac{k(k+2)}{144}
\ee
Since we had previously noted that $\mu=-\frac{c(c+4)}{576}$, it follows that $c=2k$.

Kaneko and Koike \cite{KK} considered the above equation for the following sets of values: $k=3n+\half$, $k=2n+1,n\ne 2$ mod 3, and $k=4n+2$. Thereafter, Kaneko \cite{Kaneko:On}, considered $k=\frac{6n+1}{5}, n\ne 4$ mod 5. These sets will turn out to be of crucial importance below.

From the Riemann-Roch relation $-\frac{c}{12}+h=\frac16$, one finds that 
$h=\frac{k+1}{6}$. As seen in the previous Section, $\ell=0$ characters lie in the range $-2\le c\le 10$ where the end-points are logarithmic. Correspondingly we have $-1\le k\le 5$. The rational values of $k$ that give admissible characters (including those of IVOA type) are known \cite{Mathur:1988na} to be $\frac15, \frac12, 1, \frac75, 2, \frac{13}{5},3,\frac72, \frac{19}{5}$. We did not include $k=4$ ($c=8$) because that is a one-character theory. We see that these cases precisely fall into the four categories, as follows:
\be
\begin{split}
\hbox{Lee-Yang}:&~~k=\frac{6n+1}{5},~n\ne 4~\hbox{mod}~5 ~~\to~~ k=\frac15, \frac75,\frac{13}{5},\frac{19}{5}\\
A_1:&~~k=3n+\half ~~\to~~ k=\frac12,\frac72\\
A_2:&~~k=2n+1,~n\ne 2~\hbox{mod}~3 ~~\to~~ k=1,3\\
D_4:&~~k=6n+2 ~~\to~~ k=2
\end{split}
\label{kvalues}
\ee
In the range $-1<k<5$ this is a complete set of $k$ values for admissible characters in each of the four categories. Each case labels a distinct fusion-rule class.

For values of $k$ outside this range, the above series do not give admissible RCFT characters, but remarkably they all turn out to provide quasi-characters. Starting from subsection \ref{sec.LY} we will prove this, using results of \cite{KK,Kaneko:On}, for the various different classes in \eref{kvalues}. The results can be verified by computing explicit solutions as $q$-series, several examples of which can be found in Appendix \ref{appendix.examples}. Note that the explicit formulae for characters in \eref{ellzero.hyper} continue to be valid for quasi-characters -- one simply has to substitute $c=2k$ in those formulae to get the quasi-characters at $\ell=0$ for any $k$. 

\subsubsection*{$\ell=2$ case: ``dual'' parametrisation}

While the Kaneko-Zagier equation leads, via \eref{scaledsoln}, to $\ell=0$ quasi-characters, an analogous equation leading to $\ell=2$ quasi-characters does not appear to have been studied in the literature. In fact the desired equation is:
\be
\left(D^2_{(k)}+\frac13\frac{E_6}{E_4} D_{(k)}-\frac{k(k-2)}{144}E_4(\tau)\right)f_{(k)}(\tau)=0
\label{KZdual}
\ee
We will refer to this as the ``dual Kaneko-Zagier equation''.

Scaling the solutions as in \eref{scaledsoln} to obtain weight-0 (quasi-)characters, one gets the equation:
\be
\left(D^2+\frac13\frac{E_6}{E_4} D-\frac{k(k-2)}{144}\right)\chi=0
\label{dualKZeqn}
\ee
which is a special subfamily of \eref{leq.2} obtained by setting $\mu=-\frac{k(k-2)}{144}$. We see again that the associated central charge is $c=2k$.

This time the Riemann-Roch theorem says that $h=\frac{k-1}{6}$. From Table \ref{table-leq.2}, admissible characters are bounded in a range of central charge $14\le c\le 26$, again with logarithmic theories at the end-points. This range corresponds to $7\le k\le 13$. In fact the precise values of $k$ that lead to RCFT are $\frac{41}{5}, \frac{17}{2}, 9, \frac{47}{5}, 10, \frac{53}{5}, 11,\frac{23}{2},\frac{59}{5}$ where again we have removed the single-character case $k=8$ ($c=16$). These values fall into the following four series:
\be
\begin{split}
\hbox{Lee-Yang}:&~~k=\frac{6n-1}{5},~n\ne 1~\hbox{mod}~5 ~~\to~~ k=\frac{41}{5}, \frac{47}{5},\frac{53}{5},\frac{59}{5}\\
A_1:&~~k=3n-\half ~~\to~~ k=\frac{17}{2},\frac{23}{2}\\
A_2:&~~k=2n-1,~n\ne 1~\hbox{mod}~3 ~~\to~~ k=9,11\\
D_4:&~~k=6n-2 ~~\to~~ k=10
\end{split}
\ee

As in the $\ell=0$ case, the $k$ values listed above are a complete set for the given four series in a particular range, namely $7<k<13$ (logarithmic end-points excluded). These are the admissible characters. If now we allow $k$ to take values outside this range, then we find that the corresponding solutions to \eref{dualKZeqn} are quasi-characters in every case. Examples of these too are listed in Appendix \ref{appendix.examples}. And again, the explicit solutions in \eref{elltwo.hyper} continue to be valid with the substitution $c=2k$. 

In fact we can say more: the coset pairing that was previously observed \cite{Gaberdiel:2016zke} between $\ell=0$ and $\ell=2$ characters continues to hold for these quasi-characters. It pairs an $\ell=0$ theory with a given value of $k$ with an $\ell=2$ theory with the value $\tk=12-k$. We will discuss this relation in a subsequent subsection. But first we will show, both for $\ell=0$ and $\ell=2$, that the quasi-characters can be expressed as polynomials in the characters of a particular ``base'' theory in that class. This will in particular amount to a proof of integrality of their coefficients.

\subsection{Lee-Yang series and its dual}

\label{sec.LY}

\subsubsection*{Lee-Yang series}

This series is defined by choosing the parameter $k$ in \eref{KZeqn} as follows:
\be
k=\frac{6n+1}{5},~~ n\in {\mathbb N},~ n\ne 4~ \hbox{mod }5
\label{kkaneko}
\ee
Recall that the central charge and conformal dimension associated to these characters is:
\be
c=2k=\frac{2(6n+1)}{5},~~h=\frac{k+1}{6}=\frac{n+1}{5}
\label{chkaneko}
\ee
The first case is $n=0$, which yields $c=\frac25,h=\frac15$. These are the Lee-Yang characters in the unitary presentation. As we see from the above formulae, all the remaining cases also have a central charge and dimension that is a multiple of $\frac15$. Using \cite{Christe:1988xy} one sees immediately that the fusion rules associated to them must be of Lee-Yang type. This can also be verified directly by constructing the modular transformation matrix, This explains why we referred to this series as the ``Lee-Yang class'' (with $\ell=0$). The values $n=0,1,2,3$ give genuine $\ell=0$ characters that can be found in Table \ref{table-leq.0}. For all remaining values of $n$ we find quasi-characters. 

As shown in \cite{Kaneko:On}, the solutions for this series are given recursively in terms of the characters of the Lee-Yang theory:
\be
\begin{split}
\chi^{\rm LY}_0(\tau) &= q^{-\sfrac{1}{60}}\prod_{m=0}^{\infty}\frac{1}{(1-q^{5m+1})(1-q^{5m+4})} \\
\chi^{\rm LY}_1(\tau) &= q^{\frac{11}{60}}\prod_{m=0}^{\infty}\frac{1}{(1-q^{5m+2})(1-q^{5m+3})}
\end{split}
\label{LY}
\ee
The recursion relation can be written in a (moderately) simple form using a suitable power of the quotient of the two characters, namely:
\be
t = \left(\frac{\chi_1}{\chi_0}\right)^5
\label{haupt}
\ee
Now, for all $k$ satisfying \eref{kkaneko}, the solutions are given by $\chi_0^{5k}P_{n}(t)$ and $\chi_1^{5k}P_{n}(-t^{-1})$, where $P_n(t)$ are polynomials given as follows:
\begin{align*}
\begin{tabular}[t]{|p{1cm}|p{12cm}|}
\hline
$P_0(t)$ & $1$ \\ \hline
$P_1(t)$ & $1 + 7t$\\ \hline
$P_2(t)$ & $1+ 39t-26t^2$ \\ \hline
$P_3(t)$ & $1+ 171t+ 247t^2-57t^3$ \\ \hline
$P_5(t)$ & $1 - 465t - 10385t^2 - 2945t^3 - 8370t^4 + 682t^5$ \\ \hline
$P_6(t)$ & $1 - 333t - 17390t^2 - 54390t^3 + 26640t^4 - 64158t^5 + 3774t^6$\\ \hline
$P_7(t)$ & $1 - 301t - 36421t^2 - 310245t^3 + 10535t^4 - 422303t^5 + 283843t^6 - 12857t^7$  \\ \hline
$P_8(t)$ & $1 - 294t - 101528t^2 - 1798692t^3 - 2747430t^4 - 387933t^5 - 2086028t^6 + 740544t^7 - 26999t^8$ \\ \hline 
\end{tabular}
\end{align*}
\\
Notice that since $n\ne 4$ mod 5, $P_4,P_9,P_{14}\cdots$ do not exist. For $n \geq 10$, $P_n(t)$ are given recursively in terms of $P_{n-5}(t)$ and $P_{n-10}(t)$ as follows
\begin{align}
\begin{split}
P_n(t) = \left(t^2+1\right) \left(t^4+522 t^3-10006 t^2-522 t+1\right)P_{n-5}(t) \\
 + 12\frac{(6n-29) (6 n-49)}{(n-4)(n-9)}t\left(1-11t-t^2\right)^5P_{n-10}(t)
\end{split}
\end{align}

Now it is easy to prove integrality of the coefficients. Clearly the modular function $t$ has all integer coefficients, and at each step in the recursion only a finite common denominator occurs (which in general grows with $n$). This ensures that a finite normalisation will render the character integral. 
As indicated in the discussion around \eref{idchar}, we find that the positivity behaviour of the $q$-coefficients is of two types, which were described as Type I and Type II. In the present class it turns out that $n \in \{0,1,2,3\}$ mod $10$ gives Type I quasi-characters and $n \in \{5,6,7,8\}$ mod $10$ gives Type II. In fact for the Type I case, the $q$-coefficients of the first character (suitably normalised) are all positive integers except for the first $\lfloor \frac{n}{10}\rfloor$ {\em odd} powers of $q$. For the Type II case, on the other hand,
the $q$-coefficients are all negative except for the first $\lfloor \frac{n}{10}\rfloor+1$ {\em even} powers of $q$. In both cases, the second character has all positive integer $q$-coefficients. In general the vacuum character has degeneracies, requiring an overall integer normalisation to make the coefficients of the $q$-series integral.

To derive this result, \cite{Kaneko:On} recasts the original differential equation \eref{KZeqn} as a differential equation in $t$, which (in monic form) has rational coefficients. Writing 
\be
\chi(q)=F(t)\big(\chi^{\rm LY}_0(q)\big)^{6n+1}
\label{Fdef.leq.0}
\ee
where $\chi_0^{\rm LY}$ was defined in \eref{LY}, and using \eref{haupt}, the following equation is found:
\be
F''(t)+\frac{\left( n(11t^2+66t-1)-4(t^2+11t-1)\right)}{5t(t^2+11t-1)}F'(t)+\frac{n(6n+1)(t+3)}{5t(t^2+11t-1)}F=0
\label{teqn.leq.0}
\ee
It is then verified that the polynomials $P_0\cdots P_8$ satisfy this equation (start by choosing arbitrary coefficients in each polynomial, and determine them from the equation). Finally, the recursion relation is verified using \eref{teqn.leq.0}.

\subsubsection*{Dual Lee-Yang series}

In this case we saw that the parametrisation appropriate for the Lee-Yang class is: 
\be
k = \frac{6n-1}{5},~~ n\in \mathbb{N},~~  n\neq 1 \hbox{ mod }5
\label{kdualkaneko}
\ee
With this parametrisation we have:
\be
c=2k=\frac{2(6n-1)}{5},~~h=\frac{k-1}{6}=\frac{n-1}{5}
\label{chdualkaneko}
\ee

To find the solutions of the dual Kaneko series, we consider the same variable $t$ defined in \eref{haupt}. Then we define:
\be
\chi(q)=F(t)\Big(\chi^{\rm LY}_0(q)\Big)^{6n-1}
\label{Fdef.leq.2}
\ee
Next we find that equating $F$ to any of the following polynomials $\tilde{P}_n(t)$, for the values of $n$ in \eref{kdualkaneko}, gives rise via \eref{Fdef.leq.2} to the corresponding quasi-character for that value of $n$.
\begin{align*}
\def\arraystretch{1.2}
\begin{tabular}[t]{|p{1cm}|p{12cm}|}
\hline
$\tilde{P}_2(t)$ & $1 - 66t -11t^2$\\ \hline
$\tilde{P}_3(t)$ & $1-119t+187t^2 -17t^3$ \\ \hline
$\tilde{P}_4(t)$ & $1-207 t-391 t^2-1173t^3+46t^4$ \\ \hline
$\tilde{P}_5(t)$ & $1-435 t -6670 t^2-3335 t^4 +87 t^5$ \\ \hline
$\tilde{P}_7(t)$ & $1+369 t+50594 t^2+261580 t^3+136735 t^4-151003 t^5+54858 t^6 -902 t^7$\\ \hline
$\tilde{P}_8(t)$ & $1+141 t+55037 t^2+740673 t^3+667870 t^4+932292 t^5-1640958 t^6+340374 t^7-4794 t^8$  \\ \hline
$\tilde{P}_9(t)$ & $1+53 t+80454 t^2+2277092 t^3+6441196 t^4 +4589482 t^5-6686904 t^6 +9218078 t^7 -1278731 t^8 +15847 t^9$ \\ \hline 
$\tilde{P}_{10}(t)$ & $1+162545 t^2+8777430 t^3+57609370 t^4+48470919 t^5 +29482300 t^6 -34622085 t^7 +28804685 t^8 -2925810 t^9 + 32509 t^{10}$ \\ \hline
\end{tabular}
\end{align*}
\\
For $n \geq 12$, $\tilde{P}_n(t)$ are given recursively in terms of $\tilde{P}_{n-5}(t)$ and $\tilde{P}_{n-10}(t)$ as follows:
\be
\begin{split}
\tilde{P}_n(t) = \left(t^2+1\right) \left(t^4+522 t^3-10006 t^2-522 t+1\right)\tilde{P}_{n-5}(t) \\
 + 12\frac{(6n-31) (6 n-71)}{(n-6)(n-11)}t\left(1-11t-t^2\right)^5\tilde{P}_{n-10}(t)
\end{split}
\ee
We find that these polynomials give rise to an infinite series of quasi-characters, all satisfying the dual MMS ($\ell =2$) equation. We have verified that these are of Type I for $n= 7,8,9,10 \hbox{ mod } 10$ and Type II for $n= 2,3,4,5 \hbox{ mod } 10$. Only for the values of $n = 7,8,9,10$ are these valid characters, having all positive integer $q$-coefficients. For all other values of $n$, there are coefficients of the opposite sign, hence we have quasi-characters. 

\subsection{$A_1$ series and its dual}

\subsubsection*{$A_1$ series}

This series is defined by choosing $k$ in the \eref{KZeqn} as follows \cite{KK}:
\be
k = 3n+\half,~~ n \in \mathbb{N}
\ee
With this parametrisation, we have
\be 
c = 2k = 6n+1, ~~ h = \frac{k+1}{6} = \frac{2n+1}{4}
\ee
The values $n = 0,1$ give genuine $\ell =0$ theories, which can be found in Table \ref{table-leq.0}. For all the other values of $n$, the solutions of the MLDE are quasi-characters and are given recursively in terms of the characters of the $A_{1,1}$ theory:
\be
\begin{split}
\chi_0(\tau) &= \frac{\theta_3(2\tau)}{\eta(\tau)}\\
\chi_1(\tau) &= \frac{\theta_2(2\tau)}{\eta(\tau)}
\end{split}
\ee

The quasi-characters are polynomials of the above two characters. Consider the ratio 
\be
t= \left(\frac{\chi_1}{\chi_0}\right)^4
\ee
This is related to the modular lambda function as $t(\tau) = \lambda(2\tau)$. Then the quasi-characters for all $n$ are given as $\chi_0^{6n+1}P_n(t)$ and $\chi_1^{6n+1}P_n(t^{-1})$, where the polynomials $P_n(t)$ are given as follows (this can be derived from Theorem 1(iv) of \cite{KK}):
\begin{align*}
\def\arraystretch{1.2}
\begin{tabular}[t]{|p{1cm}|p{4.5cm}|}
\hline
$P_0(t)$ & $1$ \\ \hline
$P_1(t)$ & $1 +7t$\\ \hline
$P_2(t)$ & $1-26t-39t^2$ \\ \hline
$P_3(t)$ & $1-19t- 285t^2-209t^3$ \\ \hline
\end{tabular}
\end{align*}
For $n\geq4$, 
\begin{align}
P_n(t) = (t+1) \left(t^2-34 t+1\right) P_{n-2}(t) + \frac{3 (6 n-19) (6 n-11)}{(2 n-7) (2 n-3)} t (t-1)^4 P_{n-4}(t)
\end{align}
We find that this series of quasi-characters is of type I for $n = 0,1\hbox{ mod }4$ and of type II for $n = 2,3\hbox{ mod }4$. \\

\subsubsection*{Dual $A_1$ series}

Here we propose a dual $A_1$ series of quasi-characters which satisfy the $l=2$ differential equation. This is defined by choosing $k$ in  \eref{dualKZeqn} as
\be 
k = 3n-\half, n \in \mathbb{N}
\ee
This corresponds to the following values of central charges and conformal dimensions
\be 
c = 2k = 6n-1, ~~ h = \frac{k-1}{6} = \frac{2n-1}{4}
\ee 
These quasi-characters are also expressed as polynomials of the $A_{1,1}$ theory, and for $n\geq 1$ are given by $\chi_0^{6n-1}\tilde{P}_n(t)$ and $\chi_1^{6n-1}\tilde{P}_n(t^{-1})$.

\begin{align*}
\def\arraystretch{1.2}
\begin{tabular}[t]{|p{1cm}|p{4.5cm}|}
\hline
$\tilde{P}_1(t)$ & $1-5t$ \\ \hline
$\tilde{P}_2(t)$ & $1-22t-11t^2$\\ \hline
$\tilde{P}_3(t)$ & $1+17t+187t^2 +51t^3$ \\ \hline
$\tilde{P}_4(t)$ & $1+506t^2+1288t^3+253t^4$ \\ \hline
\end{tabular}
\end{align*}
and for $n\geq5$, 
\begin{align}
\tilde{P}_n(t) = (t+1) \left(t^2-34 t+1\right) \tilde{P}_{n-2}(t) + \frac{3 (6 n-29) (6 n-13)}{(2 n-9) (2 n-5)} t (t-1)^4 \tilde{P}_{n-4}(t)
\end{align}

We find that this series of quasi-characters is of Type I for $n = 0, 3 $ mod 4 and of Type II for $n = 1, 2$ mod 4.
\subsection{$A_2$ series and its dual}

\subsubsection*{$A_2$ series}

The $A_2$ series occurs at central charges $c = 2k=4n+2, n \neq 2$ mod $3$. For $n=0$, we have the $A_{2,1}$ theory with characters
\begin{align*}
\chi_0(\tau) &= \frac{E_1^{(3)}(\tau)}{\eta(\tau)^2} \\
\chi_1(\tau) &= 3\frac{\eta(3\tau)^3}{\eta(\tau)^3}
\end{align*}
where $E_1^{(3)}$ is the weight $1$ Eisenstein series at level $3$. The quasi-characters are given as polynomials in the above characters. However, in the $A_2$ and $D_4$ classes as well as their duals, the pattern is different from the previous cases in that the second character is not obtained from the first by a simple substitution. Hence we only quote the identity character here. Using $t = \left(\frac{\chi_1}{\chi_0}\right)^3$, we have $\chi_0^{2n+1}P_n(t)$ for the identity character, where:
\begin{align*}
\def\arraystretch{1.2}
\begin{tabular}[t]{|p{1cm}|p{4cm}|}
\hline
$P_0(t)$ & $1$ \\ \hline
$P_1(t)$ & $1 +2t$\\ \hline
$P_3(t)$ & $1-14-14t^2$ \\ \hline
$P_4(t)$ & $1-12t- 60t^2-10t^3$ \\ \hline
\end{tabular}
\end{align*}
and for $n\geq 6$, we have
\begin{align}
P_{n}(t) = (1-20t-8t^2)P_{n-2}(t) + \frac{4 (2 n-9) (2 n-5)}{(n-5) (n-2)}t(1-3t-3t^2-t^3)P_{n-4}(t)
\end{align}
These are of Type I for $n = 0, 1$ mod 6 and of Type II for $n = 3, 4$ mod 6.

\subsubsection*{Dual $A_2$ series}

The $l=2$ series of $A_2$ quasi-characters occur at $c = 2k=4n-2, n\neq 1$ mod $3$. The identity character is given as $\chi_0^{2n-1}\tilde{P}_n(t)$ where:
\begin{align*}
\def\arraystretch{1.2}
\begin{tabular}[t]{|p{1cm}|p{6cm}|}
\hline
$\tilde{P}_3(t)$ & $1-10t$ \\ \hline
$\tilde{P}_5(t)$ & $1+6t+66t^2+8t^3$\\ \hline
$\tilde{P}_6(t)$ & $1+132t^2 +110t^3$ \\ \hline
$\tilde{P}_8(t)$ & $1-5t-125t^2-1225t^3-805t^4-28t^5$ \\ \hline
\end{tabular}
\end{align*}
and for $n\geq 7$,
\begin{align}
\tilde{P}_n(t) = (1-20t-8t^2)\tilde{P}_{n-2}(t) + \frac{108 (2 n-15) (2 n-7)}{(n-7) (n-4)}t(1-3t-3t^2-t^3)\tilde{P}_{n-4}(t)
\end{align}
These are of Type I for $n = 0, 5$ mod 6 and of Type II for $n = 2, 3$ mod 6.
\subsection{$D_4$ series and its dual}

\subsubsection*{$D_4$ series}

The $D_4$ series of quasi-characters occur at central charges $c = 2k=12n+4$. For $n=0$, we get the $D_{4,1}$ affine theory whose characters are
\begin{align*}
\chi_0(\tau) &= \frac{E_2^{(2)}(\tau)}{\eta(\tau)^4} \\
\chi_1(\tau) &= 8\frac{\eta(2\tau)^8}{\eta(\tau)^8}
\end{align*}
where $E_2^{(2)}$ is the weight $2$ Eisenstein series at level $2$. Again, the quasi-characters are polynomials in the above characters. Consider the ratio $t = \left(\frac{\chi_1}{\chi_0}\right)^2$, then the identity quasi-character for any $n$ is given as $\chi_0^{3n+1}P_n(t)$ with:
\begin{align*}
\def\arraystretch{1.2}
\begin{tabular}[t]{|p{1cm}|p{2cm}|}
\hline
$P_0(t)$ & $1$ \\ \hline
$P_1(t)$ & $1 -6t-3t^2$\\ \hline
\end{tabular}
\end{align*}
and for $n\geq2$,
\begin{align}
P_n(t) = (1-9t)P_{n-1}(t) + \frac{3 (3 n-7) (3 n-5)}{(2 n-5) (2 n-3)}t(t-1)^2P_{n-2}(t)
\end{align}
We find that this series of quasi-characters is of Type I for $n = 0$ mod 2 and of Type II for $n = 1$ mod 2.
\subsubsection*{Dual $D_4$ series}

Finally, the $l=2$ series of $D_4$ quasi-characters occur at central charges $c = 2k=12n-4$. The identity character is given as  $\chi_0^{3n-1}\tilde{P}_n(t)$, with:
\begin{align*}
\def\arraystretch{1.2}
\begin{tabular}[t]{|p{1cm}|p{4cm}|}
\hline
$\tilde{P}_1(t)$ & $1+15t^2$ \\ \hline
$\tilde{P}_2(t)$ & $1 -4t-10t^2-100t^3-15t^4$\\ \hline
\end{tabular}
\end{align*}
and for $n\geq 3$,
\begin{align}
\tilde{P}_n(t) = (1-9t)\tilde{P}_{n-1}(t) +\frac{192 (3 n-8) (3 n-4)}{(2 n-3) (2 n-5)} t(t-1)^2\tilde{P}_{n-2}(t)
\end{align}
We again find that this series of quasi-characters is of Type I for $n = 0$ mod 2 and of Type II for $n = 1$ mod 2.
\subsection{General approach to recursion relations}

Here we briefly describe a more general approach to recursion relations that applies for any $k$. Consider the solutions of the original Kaneko-Zagier equation \eref{KZorig}. Recall that these have modular weight $k$ and are denoted $f_{(k)}$. These solutions have the following important recursive property \cite{KK}: given a solution $f_{(k)}$ of the equation with parameter $k$, the function $f_{(k-6)}$ (for $k\neq 0,4,5$) defined as:
\be 
f_{(k-6)} = \frac{[f_{(k)},E_4]}{\Delta}
\label{recursion.k-6}
\ee
is a solution of the same equation with parameter $k-6$. Here,
\be
[f,g] = kf(D_{(l)}g)-l(D_{(k)}f)g
\ee
is the Rankin-Cohen bracket of two modular forms of weights $k$ and $l$. Also, given two solutions $f_{(k)}$ and $f_{(k-6)}$ of the equations with parameters $k$ and $k-6$ respectively, the solution of the equation with parameter $k+6i+6$ for all $i\geq 0$ is given by 
\be 
f_{(k+6i+6)} = E_6f_{(k+6i)} + 432\frac{(k+6i)(k+6i-4)}{(k+6i+1)(k+6i-5)}\Delta f_{(k+6i-6)}
\label{recursion.k+6}
\ee 

Thus, for each solution $f_{(k)}$ of the original equation, the solutions of differential equations with parameters $k-6$ and $k+6$ exist and are given in terms of $f_{(k)}$ by the simple recursive formulae Eqs.(\ref{recursion.k-6}),(\ref{recursion.k+6}).
Now from 
\cite{Mathur:1988na} we know the set of all solutions with integer coefficients for $k$ lying in the interval $[-1,5]$ (to get the weight-$k$ solution one has to multiply the solutions of \cite{Mathur:1988na} by $\eta^{2k}$). Then we can use \eref{recursion.k-6} to calculate the solutions corresponding to the interval $[-7,-1]$. Finally we can use this set in \eref{recursion.k+6} to find all the solutions for arbitrary values of $k$. This enables us to list all the quasi-characters for the $l=0$ case, starting from the characters. One can think of the range $c\in[-2,10]$ (equivalently $k\in[-1,5]$) as a ``fundamental domain'' for $\ell=0$ quasi-characters. We have noted previously that the end-points of this fundamental domain are logarithmic CFT's. The approach of this subsection will reproduce precisely the quasi-characters of the previous sections, though not directly in polynomial form.

We can carry out the same procedure for the dual equation in \eref{KZdual}. Although it is quite different from the Kaneko-Zagier equation, we have found that its solutions have a similar recursive property. Given a solution $f_{(k)}$ of \eref{dualKZeqn}, the function $f_{(k-6)}$ (for $k\neq 0,1,4$) defined as:
\be 
f_{(k-6)} = \frac{[f_{(k)},E_4]}{\Delta} 
\ee 
is a solution of the same equation with parameter $k-6$. Given the solutions $f_{(k)}, f_{(k-6)}$, we can find the solution $f_{(k+6i+6)}$ recursively, as follows:
\be 
f_{(k+6i+6)} = E_6f_{(k+6i)} + 432\frac{(k+6i)(k+6i-8)}{(k+6i-1)(k+6i-7)}\Delta f_{(k+6i-6)}
\label{recursion.dual}
\ee 
The proof of the above recursion relations closely follows that of the $l=0$ case in \cite{KK} with slight modifications. From the definition of the Rankin-Cohen bracket, we know that: 
\be 
\begin{split}
D_{(k+6)}[f_{(k)},E_4] &= D_{(k+6)}\left( -\frac{k}{3}f_{(k)}E_6 - 4D_{(k)}f_{(k)}E_4\right) \\
&= -\frac{k}{3}D_{(k)}\left(f_{(k)}\right)E_6 + \frac{k}{6}f_{(k)}E_4^2 - 4E_4D^2_{(k)}\left(f_{(k)}\right) + \frac{4}{3}E_6f_{(k)}\\
&= -\frac{k}{3}D_{(k)}\left(f_{(k)}\right)E_6 + \frac{k}{6}f_{(k)}E_4^2\\
&\qquad  - 4E_4\left( -\frac13\frac{E_6}{E_4} D_{(k)}f_{(k)}+\frac{k(k-2)}{144}\right) + \frac{4}{3}E_6f_{(k)}\\
&= \frac{(k-8)}{18}[f_{(k)},E_6]
\end{split}
\ee
Here, we have used the Ramanujan identities $D_{(4)}E_4 = -\frac{1}{3}E_6$ and $D_{(6)}E_6 = -\frac{1}{2}E_4^2$. Taking another derivative of the bracket, we can show that
\be 
D_{(k+8)}D_{(k+6)}[f_{(k)},E_4] + \frac13\frac{E_6}{E_4} D_{(k+6)}[f_{(k)},E_4] = \frac{(k-6)(k-8)}{144}E_4[f_{(k)},E_4]
\ee
Dividing the above equation by $\Delta$, and using $D_{(12)}\Delta =0$ gives us the dual KZ equation with the free parameter $k$ replaced by $k-6$, and shows that $\frac{[f_{(k)},E_4]}{\Delta}$ is its solution. The other recursion relation, is also proved in a similar manner. Note that the coefficient in front of $f_{(k+6i-6)}$ in \eref{recursion.dual} differs from the $l=0$ case. The fundamental domain for $k$ in this case will be chosen as $[7,13]$ because this gives the admissible characters. Here too, logarithmic characters lie at the boundaries of this domain.

\subsection{Modular transformations for quasi-characters}
\label{modular.quasi}

An important operation that we will consider in the following Section is the addition of quasi-characters. When suitably done, this can remove minus signs and convert quasi-characters into admissible characters. However, the process of addition must commute with modular transformations. For this to be true, the different quasi-characters being added must all have the same modular $S$ and $T$ matrices. These have been computed for admissible characters in \cite{Naculich:1988xv} in terms of trigonometric functions, and in \cite{Mathur:1988gt} in terms of $\Gamma$-functions. Because the solutions of the $\ell=0$ differential equation have universal expressions in terms of hypergeometric functions that depend only on $k$, the modular transformation matrices likewise can be expressed for all cases as a function of $k$ by simply extending these formulae to all $k$. The result can be written:
\be
S=
\begin{pmatrix}
\frac{1}{\sin \frac{\pi(k+1)}{6}}\quad
 & \left(\mathrm{M}\frac{1-4\sin^2 \frac{\pi(k+1)}{6}}{4\sin^2 \frac{\pi(k+1)}{6}}\right)^\frac12 \\[3mm]
\left(\mathrm{\frac{1}{M}}\frac{1-4\sin^2 \frac{\pi(k+1)}{6}}{4\sin^2 \frac{\pi(k+1)}{6}}\right)^\frac12\quad
 & -\frac{1}{\sin \frac{\pi(k+1)}{6}}
\\
\end{pmatrix}
\label{smatrix}
\ee
Here M is the multiplicity of the non-trivial primary. We see that the result is periodic under the shift $k\to k+12$, but not any smaller shift. This formula has a nice property under the inversion $k\to -2-k$. This shift preserves the Kaneko-Zagier parametrisation and is known to interchange the characters. One can show that under this shift, the above formula transforms by an exchange of the two rows and the two columns.

As an explicit example, for the Lee-Yang class of quasi-characters one finds:
\be
\begin{split}
n=1,2~\hbox{mod }10: & ~~S=\begin{pmatrix}\sqrt{\frac{2}{5+\sqrt5}} &
\sqrt{\frac{2}{5-\sqrt5}} \\
\sqrt{\frac{2}{5-\sqrt5}} &
-\sqrt{\frac{2}{5+\sqrt5}} \end{pmatrix}\\[2mm]
n=0,3~\hbox{mod }10: & ~~S=\begin{pmatrix}\sqrt{\frac{2}{5-\sqrt5}} &
\sqrt{\frac{2}{5+\sqrt5}} \\
\sqrt{\frac{2}{5+\sqrt5}} &
-\sqrt{\frac{2}{5-\sqrt5}} \end{pmatrix}
\end{split}\nn
\ee
\be
\begin{split}
n=6,7~\hbox{mod }10: & ~~S=\begin{pmatrix}-\sqrt{\frac{2}{5+\sqrt5}} &
\sqrt{\frac{2}{5-\sqrt5}} \\
\sqrt{\frac{2}{5-\sqrt5}} &
\sqrt{\frac{2}{5+\sqrt5}} \end{pmatrix}\\[2mm]
n=5,8~\hbox{mod }10: & ~~S=\begin{pmatrix}-\sqrt{\frac{2}{5-\sqrt5}} &
\sqrt{\frac{2}{5+\sqrt5}} \\
\sqrt{\frac{2}{5+\sqrt5}} &
\sqrt{\frac{2}{5-\sqrt5}} \end{pmatrix}
\end{split}
\label{modularclass}
\ee
The shift $k\to -k-2$ discussed above acts on the label $n$ of this case as $n\to-2-n$ and we can see that the expected pattern holds. For example if we consider $n=-10$, this will have an $S$ given by the second line in the above equation. Under inversion it becomes $n=8$ which corresponds to the fourth line. Now if we take the matrix in the fourth line and exchange the role of identity and non-identity character (which is done by exchanging the rows and then exchanging the columns) then we get back the same matrix as in the second row.

An important application of the above modular S matrix is to determine the asymptotic behaviour of the $q$-coefficients in quasi-characters. As shown in Appendix \ref{Rademacher}, this behaviour is given by \eref{asymptotic} which we reproduce here:
\be
a_j(n) \sim S^{-1}_{j0}\,a_0(0)\,e^{4\pi\sqrt{\frac{c}{24}(n+\alpha_j)}} 
\ee
Thus the asymptotic sign of the coefficients is determined by the sign of $S^{-1}_{j0}\,a_0(0)$. From this, we can immediately read off several facts about quasi-characters. Since $a_0(0)$ is always normalised to be positive, we only need to check the sign of $S^{-1}_{j0}$, i.e, the entries of the first column in modular S matrix. The Type I or Type II behaviours, in which the coefficients have asymptotically positive or negative signs respectively is determined by the sign of $S^{-1}_{00}$. From \eref{smatrix} we find:
\be
\begin{split}
0<\frac{k+1}{6}<1~~~\hbox{mod}~2:~~ & \hbox{Type I}\\
1<\frac{k+1}{6}<2~~~\hbox{mod}~2:~~ & \hbox{Type II}
\end{split}
\label{qtypes}
\ee
For example, in the Lee-Yang class, type I would occur for $n=0,1,2,3$ and type II for $n = 5,6,7,8$. Also, since $S^{-1}_{10}$ is always positive, the non-vacuum character will always be asymptotically positive.

We would also like to have an analogous result for the $\ell=2$ quasi-characters. In principle this can be deduced by the same methods as above. However there is a simpler route. In the following subsection we will demonstrate a coset relation between quasi-characters with $\ell=0$ and those with $\ell=2$. Since the coset pairing leads to a bilinear relation that is modular invariant, it follows that the modular transformation matrix for any $\ell=2$ quasi-character is the inverse of that for its $\ell=0$ coset dual. This ensures in particular the same periodicity property as above when $k\to k+12$. 

\subsection{Coset relations among quasi-characters}

If we compare the quasi-characters for $\ell=0$ and $\ell=2$, we find that
they are related as follows. Consider the former at some value of $k$ and the latter at a value of $\tk$ such that $k+\tk=12$. From Riemann-Roch it also follows that the conformal dimensions $h,\tilh$ associated to the two sets satisfy $h+\tilh=2$. Using $c=2k$, these are the same coset relations obeyed by the admissible characters at $\ell=0$ and $\ell=2$, as shown in \cite{Gaberdiel:2016zke}. Following the previous discussion, it should not be surprising that this coset (or more properly, bilinear) relation continues to hold for quasi-characters. Combining Eqs.(\ref{ellzero.hyper},\ref{elltwo.hyper}) it can be shown that:
\be
\chi_0^k\tchi_0^\tk + \chi_1^k\tchi_1^\tk=j(q)+f_k
\label{coset.quasi}
\ee
where we have denoted the $\ell=2$ quasi-characters as $\tchi_i$, and $f_k=-\frac{1728(k-4)}{2(k-5)}$. The above equation is basically the same as Eq.(4.10) of \cite{Gaberdiel:2016zke}.

It is pleasing to see that this relation, once thought to hold only for a finite set of admissible characters, holds for an infinite set of quasi-characters. One should not be surprised to see that the additive constant is, in general, a fraction. The reason is that the characters listed in Eqs.(\ref{ellzero.hyper},\ref{elltwo.hyper}) are normalised so that their $q$-series starts with 1. However, with this normalisation, quasi-characters are not always integral, but have bounded denominators. Thus upon multiplication with a suitable integer they become integral, but now must be thought of as being in the IVOA category. It is clear that the same normalisation that makes each quasi-character integral will also make the constant on the RHS of \eref{coset.quasi} integral.

\subsection{Summary of the properties of quasi-characters}

It is useful to summarise the basic properties of quasi-characters that we have discussed above. This is done in Table \ref{table-quasi}.

\begin{table}[ht]
%\TABLE{
\centering
{\renewcommand{\arraystretch}{1.4}
\begin{tabular}{|c|c|c|c|}
\hline
Class & $c$ & $h$ & $n$ values \\
\hline
LY  & $\frac25(6n+1)$ & $\frac{n+1}{5}$ & $n\ne 4$ mod 5\\
\hline
$A_1$ & $6n+1$ & $\frac{2n+1}{4}$ & $n\in {\mathbb N}$ \\
\hline
$A_2$ & $4n+2$ & $\frac{n+1}{3}$ & $n\ne 2$ mod 3 \\ 
\hline
$D_4$ & $12n+4$ & $\frac{2n+1}{2}$ & $n\in {\mathbb N}$\\
\hline
\end{tabular}
\begin{tabular}{|c|c|c|c|}
\hline
Class & $c$ & $h$ & $n$ values \\
\hline
${\widetilde {\rm LY}}$  & $\frac25(6n-1)$ & $\frac{n-1}{5}$ & $n\ne 1$ mod 5\\
\hline
${\tilde A}_1$ & $6n-1$ & $\frac{2n-1}{4}$ & $n\in {\mathbb N}$ \\
\hline
${\tilde A}_2$ & $4n-2$ & $\frac{n-1}{3}$ & $n\ne 1$ mod 3 \\ 
\hline
${\tilde D}_4$ & $12n-4$ & $\frac{2n-1}{2}$ & $n\in {\mathbb N}$\\
\hline
\end{tabular}}
\caption{Quasi-characters with $\ell=0$, $\ell=2$.}
\label{table-quasi}
%}
\end{table}

An important consequence of our discussion is that the sets of quasi-characters listed in the table, both for $\ell=0$ and $\ell=2$, are complete. This can be argued as follows. The periodicity discussed above is general for any value of $k$. Thus if the list of quasi-characters were not complete, there would be additional candidate quasi-characters in the ``fundamental domain'' $k\in [-1,5]$ for $\ell=0$ or $k\in [7,13]$ for $\ell=2$. But this cannot be the case, since we have a complete classification in this region. Another way to argue this is that the fusion-rule classification \cite{Christe:1988xy,Mathur:1989pk} for two-character theories admits conformal dimensions which can only be multiples of $\half,\frac13,\frac14,\frac15$. Now the range of $k$ considered in the various series above  exhausts all possible such values.

Notice that for $\ell=0$ the excluded values $c=\frac25(6n+1), n=4$ mod 5 as well as $c=4n+2,n=2$ mod 3 are both coincident with the series of values $c=2(6m+5)$ for some integer $m$, which we have argued describe logarithmic theories. Likewise for $\ell=2$  the excluded values $c=\frac25(6n-1), n=1$ mod 5 as well as $c=4n-2,n=1$ mod 3 are both coincident with the series $c=2(6m+1)$ for some integer $m$, which describes the logarithmic theories in this case.

Finally we note that an infinite set of $\ell=4$ quasi-characters can be obtained by simply multiplying those for $\ell=0$ by $j^\frac13$. The argument above can be extended to show that these are similarly complete. Note that multiplying $\ell=2$ quasi-characters by the same function cannot lead to a complete set for $\ell=6$ because from this case onwards, addition of quasi-characters consistent with modular invariance becomes possible and gives more general solutions.

\section{Generating characters for $\ell\ge 6$}

\label{sec.mainresult}

The main result of this Section (and of the paper) is that despite being inadmissible as RCFT characters, the quasi-characters provide useful building blocks to construct admissible characters with higher values of $\ell$. This is done by removing  the negativity in the $q$-coefficients, which can be done using two different methods as we now describe.

\subsection{Multiplicative method}

\label{subsec.products}

The first observation is that products of quasi characters of Type I with a sufficiently high degree polynomial in $\chi_{E_{8,1}}=j^\frac13$ can eliminate all of the negative signs in the former (this was correctly anticipated in \cite{Kiritsis:1988kq}). In this process, the vector-valued modular transformations of the quasi-characters are preserved up to a phase, since $j^\frac13$ is modular invariant up to a phase all by itself. It is quite evident that the minimum power of $j^\frac13$ that leads to an admissible character in this way will, by definition, give a novel theory that is not the tensor product of others. (Recall also that the only way for a two-character theory to be the tensor product of other CFT's is that it is the product of a one-character and a two-character theory.)  

It is easy to verify that if we tensor 
$j^\frac{r}{3}$ with a pair of quasi-characters having a particular value of $\ell$, the result has:
\be
\ell'=\ell+4r
\label{ellprod}
\ee
In many cases the vacuum of the quasi character is degenerate, hence the resulting set of new characters will be of ``IVOA type''. 

The first non-trival example of this kind is provided by the three admissible characters that we found with $\ell=4$ in Subsection \ref{ell.4}. These correspond to the following values of $c,h$:
\be
(c,h)=(\sfrac{162}{5},\sfrac{11}{5}),~(33,\sfrac94),~(34,\sfrac73)
\ee
It is easy to verify that each of these character pairs corresponds to the product of $\chi_{E_{8,1}}=j^{\frac13}$ with the $\ell=0$ quasi-characters having:
\be
k=\sfrac{61}{5},\sfrac{25}{2},13
\label{threequasi}
\ee
in \eref{KZeqn}. 
Notice that the $h$-values are preserved in the process: the above quasi-characters and the final admissible characters both have $h=\sfrac{11}{5},\sfrac94,\sfrac73$. Also note that the $k$-values in \eref{threequasi} belong to the series of quasi-characters $k=\frac{6n+1}{5}, n=10; k=3n+\half, n=4; k=2n+1,n=6$ respectively, which fall in the Lee-Yang, $A_1$ and $A_2$ classes respectively. From Eqs.(\ref{typeI-LY.ex}),(\ref{typeI-A1.ex}) and (\ref{typeI-A2.ex})  we see that each of these has a single negative coefficient $m_1$ at first level above the vacuum state in the identity character. Multiplication by $j^\frac13\sim q^{-\frac13}(1+248q+\cdots)$ will add 248 to this coefficient, so if $m_1 \ge -248$ then the final character after multiplication will have all positive integral coefficients and thereby become admissible. Inspection of the three examples reveals that their values of $m_1$ are $-244$, $-245$ and $-247$ respectively, satisfying the requirement. Within the space of $\ell=0$ quasi-characters, it is easy to verify that these are the only ones satisfying $m_1\ge -248$ (we do not count the admissible characters with $\ell=0$ since those give rise to tensor-product theories at $\ell=4$), which explains why there are precisely three new admissible characters with $\ell=4$.

Let us generate a few more examples involving type I quasi-characters. In the Lee-Yang class we may consider $n=20,30,40$ for which the $q$-expansions are given in \eref{typeI-LY-higher.ex}. We have verified that the following are the ``minimal'' admissible characters that can be made from these by multiplication by powers of $j^\frac13$:
\be
j^\frac43 \chi_i^{n=20},\quad j^\frac73 \chi_i^{n=30},\quad j^\frac{10}{3} \chi_i^{n=40},\quad j^\frac{13}{3} \chi_i^{n=50}
\ee
From \eref{ellprod}, these characters have $\ell=16,28,40$ and 56 respectively. In this series we see a clear pattern: quasi-characters of Lee-Yang class with $n=10m$ become admissible on multiplication by the character of $E_{8,1}$ raised to the power $3m-2$. 

In general, whenever we start with quasi-characters at $\ell=0$ we find that the products made by tensoring with $j^\frac{r}{3}$ have $\ell=4r$. If we start with dual quasi-characters at $\ell=2$, we will find theories with $\ell=4r+2$. This exhausts all even values of $\ell$.

\subsection{Additive method}

In this section we discuss how linear combinations of Type I quasi-characters can be used to create admissible characters with higher values of $\ell$. In this approach, we add different $q$-series, each corresponding to a quasi-character, such that the sum is an admissible character with only non-negative $q$-coefficients. Of course, arbitrary sums of characters destroy modular invariance so these sums need to be taken carefully. To preserve modular invariance we can only add two characters which transform in the same way under modular transformations. Moreover we should use rational coefficients when adding them, and if necessary normalise the result to integers. As we now show, the resulting theory will have the $\ell$-value increased in multiples of $6$. Moreover we will show in the following subsection that this process leads to the {\em complete} set of admissible characters for all $\ell$.

Let us start with some illuminating examples. Take the $\ell=0$ quasi-characters in the Lee-Yang family corresponding to $n=0$ and 10. From \eref{modularclass} these have the same modular transformations and so the result will also transform in the same way. Introducing an integer $N_1$, we consider the family of characters:
\be
\chi_i^{n=10}+N_1\chi_i^{n=0}
\ee 
For the identity character $i=0$, the $q$-expansion of this sum is:
\be
\begin{split}
&q^{-\frac{61}{60}}(1-244q+169641q^2+19869896q^3+\cdots)+N_1 q^{-\frac{1}{60}}(1+q+q^2+\cdots)\\
=& q^{-\frac{61}{60}}(1+(N_1-244)q+(N_1+169641)q^2+(N_1+19869896)q^3+\cdots)
\end{split}
\ee
while for $i=1$ (the non-identity character) the $q$-expansion is:
\be
\begin{split}
& q^{\frac{71}{60}}(310124 + 27523505 q+ 1012864984 q^2 + \cdots) +  
N_1 q^{\frac{11}{60}}(1+q^2+q^3+\cdots)\\
=& q^{\frac{11}{60}}(N_1+310124 q + (N_1+27523505)q^2 + (N_1+1012864984)q^3+\cdots)
\end{split}
\ee
Now if we choose $N_1\ge 244$, we have eliminated all negative signs and the resulting characters are admissible. To find out more about the potential CFT that they could describe, notice that for the identity character, the leading exponent is that of the quasi-character of higher central charge ($n=10$ in this example) while for the non-identity character the situation is reversed: the exponent is that of the quasi-character of lower central charge ($n=0$ in this example). It follows that the CFT would have $c=\frac{122}{5}$ and $h=\frac{6}{5}$, from which we find that $\ell=6$. Thus by simply adding two quasi-characters with $\ell=0$, we have found an infinite set of admissible characters with $\ell=6$, one for each integer $N_1\ge 244$. While we do not necessarily expect there to be an RCFT for each of these cases, this achieves the goal of generating large classes of admissible characters.

The next example is striking because, for some choices, it allows us to convert a quasi-character with a degenerate identity field (what we called IVOA type) to an admissible character with a non-degenerate identity. We add quasi-characters, again for the Lee-Yang series, but for the values $n=11$ and $n=1$. The result, for the identity character, is:
\be
\begin{split}
\chi_0&= q^{-\frac{67}{60}}\big(7 + (-1742 + N_1)q
+ (722729 + 14 N_1) q^2 + (133716590 + 42 N_1) q^3\\
& + (7374239425 + 140 N_1) q^4 + 
(220691372762 + 350 N_1) q^5\\
&+ (4460548657432 + 840 N_1) q^6
+ (68133599246580 + 1827 N_1) q^7+\cdots\big)
\end{split}
\ee
Like the previous case, this one again has $\ell=6$. In this case the original identity character for $n=11$ was of IVOA type, and in fact 7-fold degenerate as we see from the 7 multiplying the leading term. This means the higher degeneracies were not divisible by 7, if they had been so then we could have normalised the character to have a non-degenerate vacuum state.
However after adding characters as above, a miracle takes place when $N_1=1742$, its lowest allowed value. In this case the first level degeneracy above the identity vanishes, but all the higher degeneracies become multiples of 7. Thus we find:
\be
\begin{split}
\chi_0&=7 q^{-\frac{67}{60}} (1 + 106731 q^2 + 19112822 q^3 + 1053497615 q^4\\
& + 31527426066 q^5 + 637221445816 q^6 + 9733371775602 q^7 + \cdots)
\end{split}
\ee
We are now in a position to drop the leading 7 and find an admissible character with a non-degenerate vacuum. We see the encouraging fact that quasi-characters of IVOA type (which are the generic type) can lead to regular non-degenerate admissible characters upon being added to each other. By examining numerous examples we have found that this generically seems to happen at least for the minimal allowed value of the integer constant.

The above example shows that one should in general consider rational, rather than integer, linear combinations of quasi-characters. For example to achieve the correctly normalised admissible character in the above equation (after the overall 7 has been dropped) one would need to add $\frac17$ of one quasi-character to $\frac{N_1}{7}$ times the other one. 

In general one can take sums (with rational coefficients) of any number of quasi-characters that all have the same modular transformations and, for suitable choices of the coefficients, generate large sets of admissible characters. The question is then, what is the $\ell$-value of the result. We can provide a general formula for this. Consider a set of $\ell=0$ quasi-characters all lying in the same class (but not necessarily the Lee-Yang class). Label them by the parameter $k$. As we have seen, the shift $k\to k+12$ leaves the modular transformations invariant. Thus, we may consider sums of the form:
\be
\sum_{p=0}^{p_{\rm max}}N_p\,\chi_i^{k-12p},\quad N_0=1 
\label{sumformula}
\ee
Assuming $N_{p_{\rm max}}\ne 0$, the critical exponents of the resulting characters are:
\be
\alpha_0=-\frac{k}{12}, \quad \alpha_1=\frac{k}{12}+\frac16-p_{\rm max}
\ee
from which it follows that $\ell=6p_{\rm max}$. This agrees with our previous examples where $p_{\rm max}$ was 1 and we found $\ell=6$. Thus we see that it is no problem to generate infinite sets of admissible characters with arbitrarily large values of $\ell$ just by adding a number of quasi-characters, and choosing the integers $N_p$ to ensure that all minus signs are removed.

The above was for $\ell=0$ quasi-characters. Had we instead started with $\ell=2$ quasi-characters, we would end up with $\ell=6p_{\rm max}+2$. Finally, we have seen that all $\ell=4$ quasi-characters are products of $j^\frac13$ with $\ell=0$ quasi-characters. By adding these to each other as above, we can generate admissible characters with $\ell=6p_{\rm max}+4$. Thus we have shown how to generate infinitely many admissible characters for all even values of $\ell$.

The specific examples considered so far involved only Type I quasi-characters with a single negative coefficient. However the procedure works equally well with more negative coefficients. For example one can consider the quasi-character in the $A_1$ case at $\ell=0$ and $n=12$, see \eref{A1.neq.12}. This has negative coefficients in front of $q,q^3$ and $q^5$. On taking an arbitrary linear combination of this with the $n=8$ and $n=4$ quasi-characters in Eqs.(\ref{A1.neq.8}) and (\ref{typeI-A1.ex}) respectively, as well as the well-known $n=0$ characters (which correspond to the $SU_{2,1}$ WZW theory), we find that the identity character goes as:
\be
\begin{split}
q^{-\frac{73}{24}}\Big(&
119 + (-53363 + 13 N_1) q + (14459256 - 4361 N_1 + 
    N_2) q^2\\
    & + (-3364790387 + 1024492 N_1 - 245 N_2 + 
    N_3) q^3\\
    & + (842188593869 - 284433485 N_1 + 142640 N_2 + 
    3 N_3) q^4\\
    & + (-303881533638137 + 296843797565 N_1 + 18615395 N_2 + 
    4 N_3) q^5\\
    &+ (461207383305660887 + 84306237909803 N_1 + 837384535 N_2 + 7 N_3) q^6
     +\cdots\Big)
\end{split}
\ee
Notice that, because of the addition process, there are now minus signs at all orders from $q$ to $q^5$. However for suitable choices of the integers $N_1,N_2,N_3$ one can easily ensure that all these terms become non-negative. The subsequent terms in the above character from ${\cal O}(q^6)$ onwards are all positive linear combinations of the $N_i$, so they will remain positive if we choose all $N_i$ non-negative, and it is clear that this allows for infinitely many choices. It may further be possible to choose the $N_i$ to be rational and even negative, yet obtaining admissible characters after taking the sum. But our aim here is only to show that there are infinitely many solutions to the requirement of admissibility, and that they are easily constructed. 

In the previous subsection we saw how to get admissible characters by multiplying quasi-characters by $j^\frac{r}{3}$. One may wonder whether this approach is exhaustive, generating all admissible characters with $\ell\ge 4$. In such a situation, potentially that method would yield identical results to the one explained in this subsection. However it is easy to see that this is not the case. For example, starting from a given quasi-character with $\ell=2$, we can get a single (potentially admissible) character at $\ell=6$ upon multiplying by 
$j^\frac13$. However the methods of the present section allow for infinitely many characters, all with the same central charge and conformal dimension, at $\ell=6$. Thus the method of adding quasi-characters is more powerful.

Nonetheless the method of the previous subsection is essential. If we are interested in admissible characters with $\ell=4$ mod 6 (for example $\ell=10$) we can only get them by the addition method if we start with $\ell=4$ quasi-characters. But these, in turn, can be generated by multiplying $\ell=0$ quasi-characters by $j^\frac{1}{3}$. Thus it appears that the ``seed'' quasi-characters that could potentially generate all admissible characters, are those which we have described based on the Kaneko-Zagier parametrisation and its dual ($\ell=0$ and $\ell=2$ respectively) as well as $j^\frac13$ times the $\ell=0$ quasi-characters. Just using these three sets, one can generate infinitely many admissible characters for all even $\ell\ge 6$ by adding quasi-characters. The remarkable thing is that this process generates the complete set of admissible characters for all allowed (i.e. even) $\ell$. We will prove this below.

\subsection{Completeness of the additive method}

\label{completeness}

In this subsection we show that by adding suitable $\ell=0$ Type I quasi-characters to each other with chosen rational coefficients, one can generate {\em every} admissible character with $\ell=6m$ for all positive integers $m$. Our strategy will be to work the other way: if we are given a pair of admissible characters with $\ell=6m$, we will show that one can add quasi-characters with suitably chosen rational coefficients in such a way as to {\em reduce} the $\ell$-value to $6m-1$\footnote{We are grateful to Ashoke Sen for suggesting this strategy.}. Repeating sequentially, one is able to reduce the given pair to a linear combination of $\ell=0$ quasi-characters.

Thus let us start by considering a pair of admissible characters, assumed to be given, having $\ell=6m$ and exponents $\alpha_0=-\frac{c}{24},\alpha_1=-\frac{c}{24}+h$:
\be
\alpha_0+\alpha_1=\frac{1-\ell}{6}=\frac16-m
\ee
Hence they have expansions of the form:
\be
\begin{split}
\chi_0 &=q^{\alpha_0}(a_0^0+a_1^0q+a_2^0 q^2+\cdots)\\
\chi_1 &=q^{\alpha_1}(a_0^1+a_1^1q+a_2^1 q^2+\cdots)
\end{split}
\ee
Let us now find an $\ell=0$ pair of quasi-characters ${\hat \chi}$ in the same fusion class which, when added to the above, gives a quasi-character with the value of $\ell$ reduced by 6. From the Riemann-Roch theorem, this will happen if $\alpha_0+\alpha_1$ increases by one unit. This in turn can be done by increasing only $\alpha_0$ or only $\alpha_1$ by a single unit, or varying both such that the sum increases by one unit. Let us try to keep $\alpha_0$ fixed and increase $\alpha_1$. 

For this, we start with a pair of $\ell=0$ quasi-characters with exponents $\halpha_0,\halpha_1$ and choose $\halpha_1=\alpha_1$. It follows that $\halpha_0=\frac16-\alpha_1=\alpha_0+m$. Thus the quasi-characters are:
\be
\begin{split}
\hchi_0 &=q^{\alpha_0+m}(\ha_0^0+\ha_1^0q+\ha_2^0 q^2+\cdots)\\
\hchi_1 &=q^{\alpha_1}(\ha_0^1+\ha_1^1q+\ha_2^1 q^2+\cdots)
\end{split}
\ee
Now consider the new quasi-characters defined by:
\be
\tchi_i=\ha_0^1\,\chi_i-a_0^1\,\hchi_i
\ee
These have the following $q$-expansions:
\be
\begin{split}
\tchi_0 &=\ha_0^1\,\chi_0-a_0^1\,\hchi_0\\
&=q^{\alpha_0}\,\ha_0^1(a_0^0+a_1^0q+a_2^0 q^2+\cdots)
-q^{\alpha_0+m}\,a_0^1(\ha_0^0+\ha_1^0q+\ha_2^0 q^2+\cdots)\\
&=q^{\alpha_0}(b_0^0+b_1^0q+b_2^0q^2+\cdots)\\
\tchi_1 &=q^{\alpha_1}\Big[\ha_0^1(a_0^1+a_1^1q+a_2^1 q^2+\cdots)-
a_0^1(\ha_0^1+\ha_1^1q+\ha_2^1 q^2+\cdots)\Big]\\
&=q^{\alpha_1+1}(b_0^1+b_1^1q+b_2^1q^2+\cdots)
\end{split}
\ee
Thus the new quasi-characters $\tchi_0,\tchi_1$ have exponents $\alpha_0,\alpha_1+1$ as desired. It follows that their $\ell$-value relative to the original characters is $\tl=\ell-6$.

We can invert the relation to write the original characters as:
\be
\chi_i=\frac{1}{\ha_0^1}\Big(\tchi_i+a_0^1\hchi_i\Big)
\ee
Both objects on the RHS are quasi-characters, the first has the $\ell$-value $\ell-6$ while the second has $\ell=0$. If $\ell=6$ we are done, otherwise we can repeat the procedure to express the first term as a sum of terms with $\ell$-value $\ell-12$ and $\ell=0$. Continuing in this way we will find that our original characters are written:
\be
\chi_i=\sum_{s=0}^{s_{\rm max}} r_s\hchi_i^{(s)}=\frac{1}{A}\sum_{s=0}^{s_{\rm max}} n_s\,\hchi_i^{(s)}
\ee
where $r_s$ are rational numbers, which we have also expressed in terms of integers $A,n_s$. The characters on the RHS all have $\ell=0$.

This proves that rational linear combinations of quasi-characters generate all admissible characters. Alternatively one can take integer linear combinations up to a single overall normalisation. The same method can be easily applied to extend the theorem to admissible characters with $\ell=6m+2, 6m+4$, expressing them as rational sums of quasi-characters with $\ell=2,4$. 

\subsection{Relation to Hecke images}

In \cite{Harvey:2018rdc}, Harvey and Wu introduced novel Hecke operators that act on vector-valued modular forms which occur as RCFT characters and give rise to new sets of potential characters. These Hecke images generically have increasing values of $\ell$, and it was shown that under certain conditions they are admissible in the sense we have used in this paper. In light of the present work, we can interpret the more general Hecke images as quasi-characters. Let us briefly review their construction as it applies to the two-character case. Consider the $q$-expansions of the characters of a particular RCFT:
\be
\chi_i = \sum_{n=0}^{\infty} c_i(n) q^{n + \alpha_i}    = \sum_{m = n_i}^{\infty} b_i(m) q^{\frac{m}{N}}
\ee 
In the second expression above, $N$ is the common denominator of the original exponents $\alpha_i$, which are therefore written as $\frac{n_i}{N}$, and the summation is now over $m=n_i + nN$, with 
\be
b_i(m) = c_i(\sfrac{m-n_i}{N})
\label{bfromc}
\ee
The integer $N$ is called the conductor of the CFT. Then, the Hecke image $T_p$ of the above characters is defined to be the new $q$-series:
\be
(T_p \chi)_i =  \sum_{n=0}^{\infty} c_i^{(p)}(n) q^{n + \alpha^{(p)}_i} 		   = \sum_{m = n_i^{(p)}}^{\infty} b_i^{(p)}(m) q^{\frac{m}{N}}
\ee
Here, $p \in \mathbb{N}$ is a prime and is relatively prime to $N$. The new exponents $\alpha_i^{(p)} = \frac{n_i^{(p)}}{N}$ are defined as:
\be
\begin{split}
\alpha_0^{(p)} &= p\alpha_0\\
\alpha_1^{(p)} &= p\alpha_1~\hbox{mod 1}=p\alpha_1-\lfloor p\alpha_1\rfloor\\
\end{split}
\label{alpha.p.def}
\ee
and the new $q$-coefficients $b_i^{(p)}(m)$ are defined in terms of the old $q$-coefficients as follows:
\be 
\begin{split}
b^{(p)}_i(n) &= \begin{cases} pb_i(pn) & p\nmid n \\
                             pb_i(pn) + \rho_{ij}b_j(n/p) & p\mid n \end{cases} \\
 &= \begin{cases} pc_i(\frac{pn-n_i}{N}) & p\nmid n \\
                             pc_i(\frac{pn-n_i}{N}) + \rho_{ij}c_j(\frac{n - pn_j}{pN}) & p\mid n \end{cases}
\end{split}
\ee
for a certain matrix $\rho_{ij}$ which is completely determined by $p$. Finally the $q$-expansion can be recast in a standard form using coefficients $c_i^{(p)}$, obtained from $b_i^{(p)}$ using \eref{bfromc}.

The Hecke images $(T_p\chi)_i$ will be admissible according to our definition for certain values of $p$ which ensure that the $c_i^{(p)}$ determined by the above procedure are non-negative. For Hecke images of $\ell=0$ theories, the case predominantly considered in \cite{Harvey:2018rdc}, the $\ell$-value of the resulting characters is found by applying the Riemann-Roch theorem:
\be
\ell^{(p)}=1-6\Big(\alpha_0^{(p)}+\alpha_1^{(p)}\Big)
\label{ell.p}
\ee
Inserting \eref{alpha.p.def}, one finds:
\be
\ell^{(p)}=1+6\left\lfloor p \alpha_1\right\rfloor -p
\ee
From this one can derive that:
\be
\begin{split}
\ell^{(p)}=\hbox{0 mod 6} ~&\to~ p=\hbox{1 mod 6} \\
\ell^{(p)}=\hbox{2 mod 6} ~&\to~ p=\hbox{5 mod 6}
\end{split}
\ee
The main virtue of the Hecke image procedure is that we can easily generate infinite sets of admissible characters with arbitrarily large values of $\ell$ as well as $c$.

However this procedure has a limitation. If we want to find $\ell^{(p)}=4$ mod 6 starting from an $\ell=0$ theory, then we must have $p=3$ mod 6. But this is never realised, because $p$ has to be co-prime to the conductor $N$, and in two-character theories the conductor is always a multiple of 3. Therefore in particular one can never get an $\ell=4$ theory as a Hecke image of something with $\ell=0$. This point was noted in \cite{Harvey:2018rdc} but incorrectly attributed to the fact that \cite{Hampapura:2015cea} did not find any such theory. As we have seen in Section 
\ref{ell.4}, there are indeed admissible characters with $\ell=4$ that escaped the notice of several previous works. Yet, they are not Hecke images of $\ell=0$ objects.

This leaves open the possibility that for $\ell=0,2$ mod 6, Hecke images of a finite set of $\ell=0$ admissible characters, together with their linear combinations, span the space of quasi-characters. Let us examine this for the special case of $\ell=6$. We start by classifying all primes $p$ such that starting from an $\ell=0$ CFT, we generate $\ell=6$. Starting from the $\ell=0$ Lee-Yang CFT, the only primes $p$ that generate an admissible $\ell=6$ Hecke image are $p=61,67,73,79$\footnote{Two of these, 61 and 79, do not obey the more stringent criteria in \cite{Harvey:2018rdc} because the first has non-unitary fusion rules and the second is of IVOA type. However we have consistently included both types of cases within our broader definition of admissibility.}. If we start with the $\ell=0$ $A_{1,1}$ theory, the primes are $p=25,31$. On the $A_{2,1}$ theory one can use the prime $p=13$. Finally on $D_{4,1}$ one can use $p=7$. These exhaust all the cases leading to admissible $\ell=6$ characters using Hecke images. The central charge and conformal dimension of these characters are given in Table \ref{table-leq.6}.

\begin{table}
\centering
{\renewcommand{\arraystretch}{1.4}
\begin{tabular}{|c||c|c|c|}
\hline
No. & Description & $c$ & $h$ \\
\hline
1 & $T_{61}\,\chi_{LY}$ & $\frac{122}{5}$ & $\frac65$ \\
\hline
2 & $T_{67}\,\chi_{LY}$ & $\frac{134}{5}$ & $\frac75$ \\
\hline
3 & $T_{73}\,\chi_{LY}$ & $\frac{146}{5}$ & $\frac85$ \\
\hline
4 & $T_{79}\,\chi_{LY}$ & $\frac{158}{5}$ & $\frac95$ \\
\hline
5 & $T_{25}\,\chi_{A_{1,1}}$ & 25 & $\frac54$ \\
\hline
6 & $T_{31}\,\chi_{A_{1,1}}$ & 31 & $\frac74$ \\
\hline
7 & $T_{13}\,\chi_{A_{2,1}}$ & 26 & $\frac43$ \\
\hline
%8 & $T_{5}\chi_{E_{6,1}}$ & 30 & $\frac53$ \\
%\hline
8 & $T_{7}\,\chi_{D_{4,1}}$ & 28 & $\frac32$ \\
\hline
\end{tabular}}
\caption{Admissible Hecke images with $\ell=6$}
\label{table-leq.6}
\end{table}

We will now show that there is a sum of $\ell=0$ quasi-characters that reproduces each of these cases (this is a special case of the general theorem proved in Subsection \ref{completeness}). In fact, all one has to do is set $k=\frac{c}{2}$ for each entry in the table, and then consider the sum:
\be
\chi^k+N_1\chi^{k-12}
\ee
We have verified that for some value of $N_1$ in each case, this set precisely reproduces all the Hecke images in Table \ref{table-leq.6}. Moreover, as explained in \cite{Harvey:2018rdc}, one is allowed to consider sums of Hecke images. Likewise, we can consider varying the coefficient $N_1$. One finds that the two procedures agree. For example, for the first entry in Table \ref{table-leq.6} one has:
\be
T_{61}\,\chi_{LY}+(N_1-244)\,\chi_{LY}=\chi^{k=\frac{61}{5}}+N_1\,\chi_{LY}
\ee
where of course $\chi_{LY}$ is the same as $\chi^{k=\frac15}$.

The LHS is the sum of a Hecke image and a character, while the RHS should be seen as the sum of two quasi-characters, the second one being in fact an admissible character.
For the remaining entries of the table, one has the general result:
\be
\chi^k+N_1\chi^{k-12}=T_p\,\chi+N_1'\chi
\ee
for suitably chosen $N_1'$, where on the right-hand side, one has to adjust $p$ according to the value of $k$ (this can be read off from the table). We conclude that the space of (sums of) Hecke images at $\ell=6$ is precisely equal to the space of sums of quasi-characters. We conjecture that this equivalence, between sums of Hecke images and sums of quasi-characters, is true more generally for $\ell=0,2$ mod 6.

The relation between Hecke images and linear sums of quasi-characters, for two-character theories, is reminiscent of a well-known phenomenon in the study of Hecke images of modular functions.  If we act with a Hecke operator on $j(\tau)$, the result is a sum of $j$-functions evaluated at (shifted) multiples of $\tau$ of the form $n\tau$ and $\frac{\tau+i}{n}$. On the other hand, the result of this action can be written as a polynomial in $j(\tau)$ using meromorphy of $j$. Comparing coefficients on both sides, one finds that linear combinations of the $q$-coefficients of $j$ are equated to sums of powers of the same coefficients. For the present case, the Hecke image of a particular character $\chi$ as defined in \cite{Harvey:2018rdc} provides an analogue of the left-hand side of this relation, since its $q$-coefficients are linear combinations of the $q$-coefficients of $\chi$. Meanwhile, recall that our quasi-characters can be written as polynomials of characters (see for example the discussion in Subsection \eref{sec.LY}). Thus on the RHS we encounter powers of coefficients of the same character $\chi$. Our conjectured equivalence then becomes a nice analogue of the famous result for modular functions.

\section{Conclusions and Discussion}

In this paper we introduced the concept of quasi-characters that have all $q$-coefficients integer, but are allowed to be positive or negative. The complete family of quasi-characters for $\ell=0$ and $\ell=2$ was presented. A complete family of $\ell=4$ quasi-characters is obtained by multiplying those of $\ell=0$ by $j^\frac13$. 

Quasi-characters are naturally labelled by a rational number $k=\half c$ where $c$ would be the central charge in the admissible case. We have argued that quasi-characters of Type I (the asymptotic degeneracy is positive) can be used to generate admissible characters by addition. One takes rational linear combinations of these objects for sets of values of $k$ differing by 12, choosing the coefficients such that the result is admissible. In this way $\ell$ is increased in multiples of 6. Repeating the process for the three ``base'' sets of quasi-characters ($\ell=0,2,4$) gives us the possibility of generating admissible characters for every even $\ell$. We proved that this procedure generates all possible admissible pairs of characters.

We have conjectured that for $\ell=0,2$ mod 6, the addition of quasi-characters yields exactly the same family of admissible characters provided by taking linear combinations of Hecke images following \cite{Harvey:2018rdc}. Our procedure may be simpler because we have an explicit formula for every quasi-character, namely Eqs.(\ref{ellzero.hyper}, \ref{elltwo.hyper}, \ref{ellfour.hyper}) with $c=2k$ and the values of $k$ being read off from Table \ref{table-quasi}. These then have  to simply be inserted into the summation formula \eref{sumformula}. We noted that this postulated equivalence between Hecke images and linear combinations is an interesting analogue of well-known facts for standard Hecke operators and modular functions.

In this work we have provided a precise algorithm to classify all  admissible characters in the two-character case. This is the analogue of the well-known result for one-character RCFT, that admissible characters are classified by 
weakly holomorphic modular functions for $SL(2,\mathbb{Z})$ with non-negative coefficients and are given by suitable polynomials of $j$ multiplied by possible powers of $j^\frac13$. The next step would be to carry out a programme on the lines of \cite{Schellekens:1992db} to identify which of this large set of characters can actually describe consistent conformal field theories.

\section*{Acknowledgements}

RC acknowledges the support of an INSPIRE Scholarship for Higher Education, Government of India. We thank Ashoke Sen for a helpful suggestion on how to prove the completeness of our classification, and for discussions. 

\appendix

\Comment{

\section{Useful Identities}

\be
\begin{split}
E_2&=24\frac{\eta'}{\eta}\\
E_4&=E_2^2-12E_2'
\end{split}
\ee
we find:
\be
E_4=288\Bigg[3\left(\frac{\eta'}{\eta}\right)^2-\frac{\eta''}{\eta}\Bigg]
\ee

}

\section{Rademacher series for vector-valued modular functions}
\label{Rademacher}

In this Appendix we briefly review the generalised Rademacher series for Fourier coefficients of vector-valued modular forms. This method, for the case of the modular function $j(\tau)$ was pioneered by Rademacher \cite{Rademacher:1939fou}. It was reviewed and generalised to vector-valued modular forms in \citep{Dijkgraaf:2000, Manschot:2010mod, Cheng:2014rad}. For a vector-valued modular function, $\chi_i = q^{\alpha_i}\sum_{n=0}^{\infty}a_i(n)q^n$, the coefficients are given by
\be
a_j(n) = \sum_{i=0,1}\sum_{m+\alpha_i<0}\mathcal{K}_{n,j;m,i}\,a_i(m)
\ee
The sum on the right side involves coefficients only of the singular (or polar) part. The infinite $\times$ finite matrix $\mathcal{K}_{n,j;m,i}$ encodes the modular properties of the $\chi_i$. The above series can be written using generalised Kloosterman sums. In our case of weight zero modular functions, this reduces to: 

\begin{small}
\be 
\mathcal{K}_{n,j;m,i} = (2\pi)^2\sum_{0\le-D/C \le 1} C^{-2} e^{\frac{2\pi i}{C}\left[ (n+\alpha_j)D +(m+\alpha_i)A\right]} \,M(\gamma)^{-1}_{ji} |m+\alpha_i |\,\tilde{I}_1\left[ \frac{4\pi}{C}\sqrt{|m+\alpha_i |(n+\alpha_j)}\right]
\ee
\end{small}

Here the summation is over all coprime numbers $C,D$ and $\gamma = \begin{tiny} \begin{pmatrix}
A & B\\
C & D\\
\end{pmatrix}\end{tiny} \in SL(2,\mathbb{Z})$. The matrix $M(\gamma)$ is the modular transformation of $\chi_i$ correspoding to $\gamma$ and $\tilde{I}_1$ is a modified Bessel function. Since we want to analyse the asymptotic behaviour of $a_j(n)$, we can consider just the leading term in the series corresponding to $C=1, D=0$, and also approximate $\tilde{I}_1$ with its asymptotic expansion. In this case, $M(\gamma)$ will just be the $S$-transformation matrix of $\chi_i$. Now, the above expression reduces to 

\be
\begin{split}
a_j(n) &= \sum_{i=0,1}\sum_{m+\alpha_i <0} \left( S^{-1}_{ji}\,e^{4\pi\sqrt{|m+\alpha_i |(n+\alpha_j)}} + \cdots\right)\, a_i(m)
\end{split}
\ee
For two character theories only the vacuum character is singular, having $m+\alpha_0 <0$, while the other character always has $m+\alpha_1 >0$ (note that $m$, being the argument of the $q$-coefficient $a_i(m)$, starts from 0). Thus, the leading behaviour in the sum arises from the $i=0$ term and when $m+\alpha_0 <0$ is most negative, i.e, $m=0$ and we have:
\be
a_j(n) = S^{-1}_{j0}\,a_0(0)\,e^{4\pi\sqrt{\frac{c}{24}(n+\alpha_j)}} + \cdots
\label{asymptotic}
\ee
where we have replaced $\alpha_0$ with $-\frac{c}{24}$. In the asymptotic large $n$ limit all the corrections are subleading, and thus do not affect the sign of $a_j(n)$. Thus the asymptotic sign of the coefficients is solely determined by the sign of $S^{-1}_{j0}\,a_0(0)$. From this, the asymptotic behaviour of type I and type II quasi-characters described in \eref{qtypes} follows immediately.

\section{Examples of quasi-characters}
\label{appendix.examples}

In this Appendix we list several indicative examples of quasi-characters in all four classes: Lee-Yang, $A_1$, $A_2$ and $D_4$, for $\ell=0$ and 2. These examples help us identify patterns in the occurrence of positive and negative signs. In every case $c=2k$, as explained in the main body of the paper, and $k$ is given by a different formula in terms of an integer $n$ for every class. These examples are restricted up to order $q^8$ in every case for brevity of presentation only. We have in fact examined them to much higher orders, and in particular have verified that in all the given examples, no sign changes occur beyond the order quoted.

\subsection{Lee-Yang class, $\ell=0$}

Here $k=\sfrac{6n+1}{5}$, $n\ne 4$ mod 5.

\subsubsection*{\underline{$\mathbf n=0,1,2,3$ mod 10 (Type I)}}

The cases $n=0,1,2,3$ correspond to admissible characters and were already described in the text. For $n=10$, corresponding to $c=\frac{122}{5}$, the $q$-expansion of the quasi-characters behaves like:
\be
\begin{split}
\chi_0 &= q^{-\frac{61}{60}}(1-244 q+169641 q^2+19869896 q^3+835603132 q^4+20272831988 q^5\\
&\qquad\qquad +343661522389 q^6+4500821047844 q^7+48387640292495 q^8+\cdots)\\
\chi_1 &= q^{\frac{71}{60}} (310124+27523505 q+1012864984 q^2+22618816409 q^3+361844239824 q^4\\ 
&\qquad\qquad +4537297743420 q^5+47139386040008 q^6+421093613229509 q^7\\
&\qquad\qquad +3320308929521868 q^8+\cdots)
\end{split}
\label{typeI-LY.ex}
\ee
Notice that $\chi_0$ has one negative coefficient, at level 1, while $\chi_1$ is entirely positive. We find that the same feature holds for $n=11,12,13$.

Next we list the identity quasi-character for $n=20,30,40$:
\be
\begin{split}
n&=20:\\
\chi_0 &=q^{-\frac{121}{60}} (4-1331 q+312224 q^2-88167211 q^3
+112365043648 q^4\\
&\qquad \qquad +29255027806796 q^5+2928616139511040 q^6+169826714738872785 q^7\\
&\qquad\qquad +6732306371343381004 q^8+\cdots)
\end{split}
\ee
\be
\begin{split}
n&=30:\\
\chi_0 &=q^{-\frac{181}{60}}(13-5792 q+1562573 q^2-363356957 q^3+91523408571 q^4\\
&\qquad \qquad -33856651489245 q^5+63239825698520008 q^6\\
&\qquad\qquad +25767972729505769158 q^7 +4186462633071570762411 q^8+\cdots)
\end{split}
\ee
\be
\begin{split}
n&=40:\\
\chi_0&= q^{-\frac{241}{60}}(93-52297 q+16786373 q^2-4226980217 q^3+979743639746 q^4\\
&\qquad \qquad -237470167054347 q^5+69820655915636751 q^6\\
&\qquad\qquad -32441367481781802004 q^7
+79958952250135225220894 q^8+\cdots)
\end{split}
\label{typeI-LY-higher.ex}
\ee
We see that negative coefficients occur at levels 1,3 for $n=20$, at levels 1,3,5 for $n=30$ and at levels 1,3,5,7 for $n=40$. From what we have said previously, the properties for $n=10p$ also carry over identically to $n=10p+1,10p+2,10p+3$.

\subsubsection*{\underline{$n=5,6,7,8$ mod 10 (Type II)}}

For $n=5$, corresponding to $c=\frac{62}{5}$, the $q$-expansion of the quasi-characters is:
\be
\begin{split}
\chi_0&= q^{-\frac{31}{60}}(1 - 434 q - 21979 q^2 - 381114 q^3 - 4097983 q^4 - 32826830 q^5\\
&\qquad\qquad - 214422474 q^6 - 1202009624 q^7 - 5975933643 q^8 - \cdots)\\
\chi_1 &= q^{\frac{41}{60}} (682 + 25420 q + 390197 q^2 + 3917222 q^3 + 29953657 q^4 + 
  189141446 q^5\\
  &\qquad\qquad  + 1032826039 q^6 + 5027747958 q^7 + 22278884080 q^8 + \cdots)
\end{split}
\label{typeII-LY.ex}
\ee

Here $\chi_0$ has one positive coefficient, at level 0, and the remaining coefficients are all negative integers. Meanwhile $\chi_1$ is still positive. We find that the same features hold for $n=6,7,8$. 

Next we list the identity quasi-character for $n=15,25,35$:
\be
\begin{split}
n&=15:\\
\chi_0 &=q^{-\frac{91}{60}}(11-3094 q+767221 q^2-752877034q^3-141459978933 q^4\\
&\qquad \qquad -9933052272234 q^5-400991441361238 q^6-11102490443045248 q^7\\
&\qquad\qquad -232430316159124902 q^8-\cdots)
\end{split}
\ee
\be
\begin{split}
n&=25:\\
\chi_0 &=  q^{-\frac{151}{60}}(14-5436 q+1349789 q^2-319951786 q^3+103851231953 q^4\\
&\qquad\qquad -163094441135962 q^5 -54419561760181373 q^6-\\
&\qquad \qquad 7127842239228463298 q^7
 -545687937386371719084 q^8-\cdots)
\end{split}
\ee
\be
\begin{split}
n&=35:\\
\chi_0 &=q^{-\frac{211}{60}}(403-202982 q+59785373 q^2-14330896762 q^3+3353464036546 q^4\\
&\qquad\qquad -911150391733816 q^5+379798015344511387 q^6\\
&\qquad\qquad -822664458350625757698 q^7
-396163375086525572637995 q^8-\cdots)
\end{split}
\ee
In these cases of type II, positive coefficients occur at levels 0,2 for $n=15$, at levels 0,2,4 for $n=25$, and at levels 0,2,4,6 for $n=35$. As we have said previously, the properties for $n=10p+5$ also carry over identically to $n=10p+6,10p+7,10p+8$.

We see that type I and type II solutions in this family have distinctive signatures of how their positive and negative signs occur, as expected from the modular behaviour. The pattern is that for the identity character, the first $\lfloor \frac{n}{10}\rfloor$ odd-level coefficients are negative in the type I case, while the first $\lfloor \frac{n}{10}\rfloor+1$ even-level coefficients are positive in the type II case. Meanwhile the second character has all positive integral coefficients. 

\subsection{Dual Lee-Yang class: $\ell=2$}

Here $k=\frac{6n-1}{5}$, $n \ne 1$ mod 5. 

\subsubsection*{\underline{$n = 7,8,9,0 \hbox{ {\rm mod} } 10$ (type I)}}

Here $n=7,8,9,10$ corresponds to admissible characters. The remaining cases of this kind are all Type I quasi-characters. As a first example,
for $n =20$, corresponding to $c = \frac{238}{5}$, the $q$-expansion is as follows:
\be
\begin{split}
\chi_0 & = q^{-\frac{119}{60}}(1 - 136 q - 7905 q^2 + 16416220 q^3 + 31847544099 q^4 \\
&\qquad\qquad +6795085533988 q^5 + 628571364597532 q^6 + 34547890520944536 q^7\\
&\qquad\qquad  + 
   1311257024182478785 q^8 + \cdots )\\
\chi_1 &= q^{\frac{109}{60}} (5726299516 + 1560416618110 q + 164844563835924 q^2 \\
&\qquad\qquad+ 9904949268352194 q^3 + 401446661502132820 q^4\\
&\qquad\qquad  + 12094697551326910394 q^5 + 288285213735532329520 q^6\\
&\qquad\qquad +5675240014977817967176 q^7 + 95198705239902059605416 q^8 + \cdots)
\end{split}
\ee  

Here, $\chi_0$ has two negative coefficients at levels 1 and 2, while $\chi_1$ is all positive. The same feature holds for $n = 17,18,19$. As $n$ is increased, the number of negative coefficients increases in $\chi_0$, while $\chi_1$ is always positive. Note that unlike the previous cases, the negative signs do not always alternate. 

Examples of the identity character are given below for $n=30,40,50$:
\be
\begin{split}
n&= 30: \\
\chi_0 &= q^{-\frac{179}{60}}(38-10203 q+1172450 q^2+67967195 q^3-109671753798 q^4\\
&\qquad\qquad +98221874676687 q^5+236681250333480608 q^6\\
&\qquad\qquad+75431777433505457548 q^7 
 +11185497932632172427260 q^8 + \cdots)
\end{split}
\ee
\be
\begin{split}
n&= 40:\\
\chi_0 &= q^{-\frac{239}{60}}(493-194068 q+38451515 q^2-4194122570 q^3-242912003863 q^4\\
&\qquad\qquad +364642005616530 q^5 -218068693037881270 q^6\\
&\qquad\qquad +201481497955252539938 q^7+596568299271483093895235 q^8 + \cdots)
\end{split}
\ee
\be
\begin{split}
n&= 50:\\
\chi_0 &= q^{-\frac{598}{5}}(5423-2800733 q+749451205 q^2-132408241035 q^3\\
&\qquad\qquad +14114193629547 q^4 +817113659718383 q^5\\
&\qquad\qquad -1189034266534646178 q^6+ 617639990945491239758 q^7\\
&\qquad\qquad -348610492187897163472770 q^8+\cdots)
\end{split}
\ee

\subsubsection*{\underline{$n = 2,3,4,5$ \hbox{{\rm mod }}10 (type II)}}

For the case of $n=15$, corresponding to $c= \frac{178}{5}$, the $q$-expansions are as follows
\be 
\begin{split}
\chi_0 &= q^{-\frac{89}{60}}(3 - 178 q -117035 q^2 - 215987870 q^3 - 34326866748 q^4 \\
 &\qquad\qquad - 2230125889140 q^5  - 84929896858225 q^6 - 2239096669123902 q^7 \\ 
 &\qquad\qquad -44900813135469340 q^8 \cdots)\\
\chi_1 &= q^{\frac{79}{60}} (40506214 + 8421028700 q + 628882423316 q^2 + 26206218083976 q^3 \\
 &\qquad\qquad+  736919256033870 q^4 + 15521985292325428 q^5\\
 &\qquad\qquad +  261309104707682010 q^6 + 3672723060854041492 q^7\\
 &\qquad\qquad +  44455408845153799482 q^8 + \cdots)
\end{split}
\ee
\noindent
The cases of $n =12,13,14$ are similar, with all negative coefficients in $\chi_0$, while $\chi_1$ remaining positive. 

As $n$ increases, the number of positive signs increases in $\chi_0$, as can be seen in the following examples:
\be
\begin{split}
n&= 25: \\
\chi_0 &=q^{-\frac{149}{60}}(19-3874 q+213815 q^2+122840070 q^3-124228171284 q^4\\
&\qquad\qquad -267787826986436 q^5 -71350964443224149 q^6\\
&\qquad\qquad -8587600018833259326 q^7 -623564032958632850035 q^8 - \cdots)
\end{split}
\ee
\be
\begin{split}
n&=35:\\
\chi_0 &=q^{-\frac{209}{60}}(29-9614 q+1534060 q^2-=83113800 q^3-46246756149 q^4\\
&\qquad\qquad +33325629324218 q^5 -29723660952695853 q^6\\
&\qquad\qquad -79681996220033722306 q^7 -29516739316725781789515 q^8 - \cdots)
\end{split}
\ee
\be
\begin{split}
n&=45:\\
\chi_0 &=q^{-\frac{269}{60}}(6409-2917574 q+681846405 q^2-99899345070 q^3\\
&\qquad\qquad +5372523175666 q^4 +2952275928321152 q^5\\
&\qquad\qquad -1907584898687122697 q^6 +1080919791498590746274 q^7\\
&\qquad\qquad -1050486606203669559554785 q^8 - \cdots)
\end{split}
\ee

As in the Lee-Yang series, the dual Lee-Yang series exhibits a pattern of signs, though in this case the negative signs do not necessarily alternate.

\subsection{$A_1$ class, $\ell=0$}

Here $k=3n+\half$ with $n\in {\mathbb Z}$.

\subsubsection*{\underline{$n = 0,1$ {\rm mod} $4$ (type I)}}

The values $n=0,1$ correspond to admissible characters, in fact they describe the $A_{1,1}$ and $E_{7,1}$ WZW models.

For the case of $n=4$, corresponding to $c= 25$, the $q$-expansions of the characters are as follows
\be 
\begin{split}
n&=4\\
\chi_0 &= q^{-\frac{25}{24}}(1-245 q+142640 q^2+18615395 q^3+837384535 q^4+21412578125 q^5\\
&\qquad\qquad +379389640345 q^6+5165089068645 q^7+57498950829715 q^8+\cdots)\\
\chi_1 &= q^{\frac{29}{24}}(1105+101065 q+3838295 q^2+88358360 q^3+1454696521 q^4\\
&\qquad\qquad+18742858160 q^5 +199800669415 q^6\\
&\qquad\qquad +1829051591175 q^7+14763322626790 q^8+\cdots)
\end{split}
\label{typeI-A1.ex}
\ee

Here, only the first level coefficient in $\chi_0$ is negative, and the rest are positive. For higher values of $n$, there are more negative coefficients. For a given $n$ in type I, there are precisely $\lfloor\frac{n}{4}\rfloor$ negative values at the odd levels in the vacuum characters. This is illustrated in the examples below for $n=8,12,16$ and the behaviour is similar for all $n = 4p, 4p+1$.

\be 
\begin{split}
n&=8 :\\
\chi_0 &= q^{-\frac{49}{24}}(13-4361 q+1024492 q^2-284433485 q^3+296843797565 q^4\\
&\qquad\qquad +84306237909803 q^5+8867059968079425 q^6+534104386666020723 q^7\\
&\qquad\qquad +21861967373053966060 q^8+\cdots)
\end{split}
\label{A1.neq.8}
\ee
\be
\begin{split}
n&=12 :\\
\chi_0 &= q^{-\frac{73}{24}}(119-53363 q+14459256 q^2-3364790387 q^3+842188593869 q^4\\
&\qquad\qquad -303881533638137 q^5+461207383305660887 q^6\\
&\qquad\qquad+203501875932273013375 q^7 +34505840401212977669887 q^8+\cdots)
\end{split}
\label{A1.neq.12}
\ee
\be
\begin{split}
n&= 16 :\\
\chi_0 &= q^{-\frac{97}{24}}(145-81965 q+26420084 q^2-6670923673 q^3+1546888407781 q^4\\
&\qquad\qquad -373699489005685 q^5+108738012982978045 q^6\\
&\qquad\qquad-49082940350350362621 q^7 +97914872285239426268650 q^8+\cdots)
\end{split}
\ee

\subsubsection*{\underline{$n = 2,3$ {\rm mod } $4$ (type II)}}

The case of $n=2$ and $c=13$, falls into type II, where all the coefficients except for the vacuum are negative:
\be 
\begin{split}
n&=2\\
\chi_0 &= q^{-\frac{13}{24}}(1-377 q-22126 q^2-422123 q^3-4875624 q^4-41490618 q^5\\
&\qquad\qquad -285717887 q^6-1679791880 q^7-8723632242 q^8-\cdots)\\
\chi_1 &= q^{\frac{17}{24}}(39+1547 q+25211 q^2+266578 q^3+2136914 q^4+14088945 q^5\\
&\qquad\qquad +80086188 q^6+404806883 q^7+1858688156 q^8+ \cdots)
\end{split}
\ee
For higher values of $n$, we encounter more positive signs in the vacuum character, precisely $\lfloor \frac{n}{4} \rfloor +1$ of them at the even levels. This is illustrated below for the cases of $n=6,10,14$:
\be
\begin{split}
n&=6:\\
\chi_0 &= q^{-\frac{37}{24}}(3-851 q+209050 q^2-169280365 q^3-34946225090 q^4\\
&\qquad\qquad -2594475493656 q^5-109380991074777 q^6-3141479177618210 q^7\\
&\qquad\qquad -67917756796632125 q^8-\cdots)
\end{split}
\ee
\be
\begin{split}
n&=10:\\
\chi_0 &= q^{-\frac{61}{24}}(221-86437 q+21544834 q^2-5090957703 q^3+1616968410460 q^4\\
&\qquad\qquad -2069802001023406 q^5-750078047818898331 q^6\\
&\qquad\qquad-102812695519515951892 q^7 -8147214212409700602408 q^8-\cdots)
\end{split}
\ee
\be
\begin{split}
n&=14:\\
\chi_0 &= q^{-\frac{85}{24}}(5-2533 q+749326 q^2-180005107 q^3+42056331894 q^4\\
&\qquad\qquad -11328519230340 q^5+4595362190935913 q^6\\
&\qquad\qquad -8069868314474265562 q^7 -4199337514335816874209 q^8-\cdots)
\end{split}
\ee

\subsection{$A_1$ class, $\ell=2$}

Here $k=3n-\half$ with $n\in{\mathbb Z}$.

\subsubsection*{\underline{$n = 0,3$ {\rm mod} $4$ (type I)}}

Here $n=3,4$ correspond to the coset duals of the $E_{7,1}$ and $A_{1,1}$ models respectively \cite{Gaberdiel:2016zke}. For $n=7$ ($c= 41$) the quasi-characters are given below. Here we see that there are two negative signs in the vacuum character:
\be 
\begin{split}
n &= 7:\\
\chi_0 &= q^{-\frac{41}{24}}(3-287 q-63304 q^2+159877409 q^3+56364700185 q^4\\
&\qquad\qquad +6045179171835 q^5 +340384709166307 q^6+12471532745472243 q^7\\
&\qquad\qquad +333677615118993343 q^8+\cdots)\\
\chi_1 &= q^{\frac{37}{24}}(101065+23976185 q+2123762075 q^2+105838183428 q^3\\
&\qquad\qquad+3554218649463 q^4 +88989323742226 q^5 +1770817971299169 q^6\\
&\qquad\qquad+29251787099757535 q^7 +413852859565527255 q^8+\cdots)
\end{split}
\ee 

As $n$ is increased, the number of negative signs increases. In this series we seem to find that the number of negative signs is always even, as can be seen in the following examples:
\be 
\begin{split}
n&= 12:\\
\chi_0 &= q^{-\frac{71}{24}}(19-5041 q+566651 q^2+38160654 q^3-56499412980 q^4\\
&\qquad\qquad +52579037361787 q^5+99364290811035618 q^6\\
&\qquad\qquad+29871659956022280223 q^7 +4261546679660911313493 q^8+\cdots)
\end{split}
\ee
\be
\begin{split}
n&= 16:\\
\chi_0 &= q^{-\frac{95}{24}}(207-80845 q+15869275 q^2-1696166150 q^3-114091964970 q^4\\
&\qquad\qquad +156251968750821 q^5-94027345456493560 q^6\\
&\qquad\qquad +91205434114115358075 q^7 +213053501796912535951850 q^8+\cdots)
\end{split}
\ee
\be
\begin{split}
n&= 20:\\
\chi_0 &= q^{\frac{119}{24}}(10695-5490813 q+1460144535 q^2-255920115650 q^3\\
&\qquad\qquad+26754714308000 q^4 +1798681240371035 q^5\\
&\qquad\qquad-2382791816599607114 q^6+1236015561687952418675 q^7\\ &\qquad\qquad-707972941827895495970625 q^8+\cdots)
\end{split}
\ee 
\subsubsection*{\underline{$n = 1,2$ {\rm mod} $4$ (type II)}}

The type II examples in the dual $A_1$ series occur for $n = 1,2$ mod $4$. For $n=1 (c=5)$, the characters are as follows
\be 
\begin{split}
n&= 1:\\
\chi_0 &= q^{-\frac{5}{24}}(1 - 65 q - 450 q^2 - 2175 q^3 - 8510 q^4 - 28672 q^5 - 86915 q^6 -
 242750 q^7\\
 &\qquad\qquad - 634875 q^8 - \cdots)\\
\chi_1 &= q^{\frac{1}{24}}(5+49 q+345 q^2+1550 q^3+5850 q^4+19055 q^5+56624 q^6+155145 q^7\\
 &\qquad\qquad +400250 q^8+\cdots)
\end{split}
\ee 
 
This time the number of positive signs seems to be odd. As usual it increases for higher values of $n$, as seen in the following examples: 
\be 
\begin{split}
n&= 5:\\
\chi_0 &= q^{-\frac{29}{24}}(5 - 29 q - 656270 q^2 - 177609775 q^3 - 12599986700 q^4\\
&\qquad\qquad -  456878461590 q^5 - 10830369716387 q^6 - 190029360086300 q^7\\
 &\qquad\qquad -2655330347867300 q^8-\cdots)
\end{split}
\ee
\be
\begin{split}
n&= 9:\\
\chi_0 &= q^{-\frac{53}{24}}(13 - 2173 q+ 12190 q^2 + 133562597 q^3 - 284193480738 q^4\\
&\qquad\qquad - 
 123295059352316 q^5 - 17683851632519551 q^6\\
 &\qquad\qquad - 1367597195527452370 q^7 - 69235287860164540405 q^8-\cdots)
  \end{split}
\ee
\be
\begin{split}
n& = 13:\\
\chi_0 &= q^{-\frac{77}{24} }(17 - 5049 q + 687170 q^2 - 3826075 q^3 - 37923365480 q^4\\
&\qquad\qquad +  27798273601142 q^5 - 63970383108405455 q^6\\
&\qquad\qquad -  38088966597755021960 q^7 - 8187834816976242242650 q^8-\cdots)
\end{split}
\ee 

\subsection{$A_2$ class, $\ell=0$}

Here $k=2n+1$ with $n\ne 2$ mod 3.

\subsubsection*{\underline{$n = 0,1$ {\rm mod} $6$ (type I)}}
For this class, we have type I behaviour when $n = 0,1\hbox{ mod }6$. The $A_{2,1}$ and $E_{6,1}$ theories correspond to $n =0,1$ and fall in this class. For $n = 6$, we have the following set of quasi-characters with one negative entry at level one.
\be 
\begin{split}
n&=6:\\
\chi_0 &= q^{-\frac{13}{24}}(1-247 q+116129 q^2+18301257 q^3+921566178 q^4+25718767594 q^5\\
&\qquad\qquad +490350941379 q^6+7117968193843 q^7+83938299826212 q^8+\cdots)\\
\chi_1 &= q^{\frac{5}{14}}(13+1248 q+49869 q^2+1205867 q^3+20800611 q^4+280099755 q^5\\
&\qquad\qquad +3113667440 q^6+29664423126 q^7+248750927451 q^8+\cdots)
\label{typeI-A2.ex}
\end{split}
\ee
As $n$ is increased in this range, the negative entries alternate, occurring only at the odd levels. For a given $n$, there are precisely $\lfloor\frac{n}{6}\rfloor$ of them:
\be 
\begin{split}
n&= 12:\\
\chi_0 &= q^{-\frac{25}{12}}(1-340 q+80105 q^2-21682360 q^3+17634482620 q^4\\
&\qquad\qquad+5846038644500 q^5 +668605708886165 q^6\\
&\qquad\qquad+42923909261295140 q^7+1853480375280258255 q^8+\cdots)\\
n&=18:\\
\chi_0 &= q^{-\frac{37}{12}}(52-23569 q+6432968 q^2-1499192825 q^3+371604861680 q^4\\
&\qquad\qquad-129048719603388 q^5+150716666353986200 q^6\\
&\qquad\qquad+76703504401244920667 q^7+13988895420513457657460 q^8+\cdots)\\
n&= 24:\\
\chi_0 &= q^{-\frac{49}{12}}(209 - 119168 q + 38681041 q^2 - 9811365207 q^3 + 2277101875209 q^4 \\
&\qquad\qquad -   547287474661207 q^5 + 156658488649829082 q^6\\
&\qquad\qquad -   67610903724358072750 q^7  + 103126878462658278760707 q^8 +\cdots)
\end{split}
\ee
\subsubsection*{\underline{$n = 3,4$ {\rm mod} $6$ (type II)}}
For the value $n = 3$ ($c= 14$), only the vacuum entry is positive:
\be 
\begin{split}
n&=3:\\
\chi_0 &= q^{-\frac{7}{12}}(1 - 322 q - 24241 q^2 - 541534 q^3 - 7036757 q^4 - 66103954 q^5\\
&\qquad\qquad -  496234060 q^6 - 3152862124 q^7 - 17578786722 q^8-\cdots)\\
\chi_1 &= q^{\frac{3}{4}}(7+306 q+5481 q^2+62958 q^3+544194 q^4+3845016 q^5\\
&\qquad\qquad +23310567 q^6+125168022 q^7+608508153 q^8+\cdots)
\end{split}
\ee
For higher values of $n$, there are $\lfloor\frac{n}{6}\rfloor + 1$ positive signs at even levels, including the vacuum entry.
\be 
\begin{split}
n&=9:\\
\chi_0 &=q^{-\frac{19}{12}}(7 - 2014 q + 488243 q^2 - 311930486 q^3 - 76043869993 q^4\\
&\qquad\qquad -  6202403199426 q^5 - 281157924188134 q^6\\
&\qquad\qquad - 8583041575115560 q^7 - 195761991550130269 q^8-\cdots)
\end{split}
\ee
\be
\begin{split}
n&=15:\\
\chi_0 &= q^{-\frac{31}{12}}(13 - 5146 q + 1290995 q^2 - 303673830 q^3 + 93315484365 q^4 \\
&\qquad\qquad - 92386057658726q^5  - 38802289774810842 q^6\\
&\qquad\qquad -5746129347161728640 q^7 - 482524744679511189450 q^8- \cdots)
\end{split}
\ee
\be
\begin{split}
n&=21:\\
\chi_0 &= q^{-\frac{43}{12}}(38 - 19436 q + 5791627 q^2 - 1396413734 q^3 + 325517062535 q^4\\
&\qquad\qquad   - 86505247171718 q^5 + 33645934235543257 q^6\\
&\qquad\qquad -  45301120823031339926 q^7 - 27095462854127103760224 q^8-\cdots) 
\end{split}
\ee

\subsection{$A_2$ class, $\ell=2$}
Here $k = 2n-1$ with $n \neq 1\hbox{ mod } 3$
\subsubsection*{\underline{$n = 0,5$ {\rm mod} $6$ (type I)}}
We have type I behaviour of the coefficients for $n=0,5\hbox{ mod }6$, where $n = 5,6$ correspond to the coset duals of $A_{2,1}$ and  $E_{6,1}$ theories \cite{Gaberdiel:2016zke}. For $n=11$ ($c=42$) we there are two negative signs in the vacuum characters:
\be 
\begin{split}
n&=11:\\
\chi_0& = q^{-\frac{7}{4}}(1-102 q-18765 q^2+39795950 q^3+16740216360 q^4+1978900454556 q^5\\
&\qquad\qquad +119922178727975 q^6+4670316757961370 q^7\\
&\qquad\qquad+131761565175644100 q^8+\cdots)\\
\chi_1 &= q^{\frac{19}{12}}(26 + 6308 q + 575035 q^2 + 29540350 q^3 + 1022588135 q^4 + 
  26376065902 q^5 \\
  &\qquad\qquad + 540255185558 q^6 + 9177898529280 q^7 +  133419906570575 q^8 +\cdots)
\end{split}
\ee

Below we list the vacuum quasi-characters for higher values of $n$ in this series. As $n$ increases, we have more negative signs but unlike in previous examples of Type I, they do not alternate with even signs:
\be 
\begin{split}
n&=17:\\
\chi_0 &= q^{-\frac{11}{4}}(13-3102 q+279081 q^2+50102074 q^3-50973132540 q^4\\
&\qquad\qquad +86221671859620 q^5+51664145678374757 q^6\\
&\qquad\qquad +10030512140941003998 q^7+1051763462636893412373 q^8+\cdots)
\end{split}
\ee
\be
\begin{split}
n&=23:\\
\chi_0 &= q^{-\frac{15}{4}}(38-13860 q+2503575 q^2-217281450 q^3-38724843105 q^4\\
&\qquad\qquad+34045263974574 q^5 -22555183763020950 q^6\\
&\qquad\qquad+43103803430683690200 q^7+33569193336853134282975 q^8+\cdots)
\end{split}
\ee
\be
\begin{split}
n&= 29:\\
\chi_0 &= q^{-\frac{19}{4}}(275-134178 q+33804135 q^2-5516807450 q^3+471561272325 q^4\\
&\qquad\qquad -69594770512954544 q^6+36491781286887210900 q^7\\
&\qquad\qquad-24537479677596020190825 q^8+\cdots)
\end{split}
\ee

\subsubsection*{\underline{$n = 2,3$ {\rm mod} $6$ (type II)}}

For the case of $n = 8$ ($c = 30$), we have the following set of quasi-characters:
\be 
\begin{split}
n&= 8:\\
\chi_0 &= q^{-\frac{5}{4}}(1 - 15 q - 98325 q^2 -32772425 q^3 - 2612032635 q^4 -
 103365035847 q^5\\
 &\qquad\qquad -2633413278190 q^6 - 49181248094250 q^7 -726558198198750 q^8-\cdots)\\
\chi_1 &= q^{\frac{13}{12}}(5+884 q+53395 q^2+1788675 q^3+40641850 q^4+697220705 q^5\\
&\qquad\qquad +9639433801 q^6+112149321915 q^7+1131934954100 q^8+\cdots)
\end{split}
\ee

\subsection{$D_4$ class, $\ell=0$}

Here $k = 6n+2$, $n \in \mathbb{N}$.

\subsubsection*{\underline{$n = 0$ {\rm mod} $2$ (type I)}}

For even values of $n$ in this series, the quasi-characters show type I behaviour. We can see that for $n =0$ we have $c=4$, corresponding to the $D_{4,1}$ theory. For the next few values of $n$ in this series, the $q$-series of characters are as follows:

\be 
\begin{split}
n&=2:\\
\chi_0 &= q^{-\frac{7}{6}}(1-252 q+90874 q^2+21857752 q^3+1383626251 q^4+45982062532 q^5\\
&\qquad\qquad +1013991773438 q^6+16709119013720 q^7+220747101927933 q^8+\cdots)\\
\chi_1 &= q^{\frac{4}{3}}(7+736 q+32368 q^2+859264 q^3+16200632 q^4+237387584 q^5\\
&\qquad\qquad +2859708096 q^6+29416423680 q^7+265458970876 q^8+\cdots)
\end{split}
\ee

\be 
\begin{split}
n&=4:\\
\chi_0 &= q^{-\frac{13}{6}}(7-2444 q+580034 q^2-150594808 q^3+87983521171 q^4\\
&\qquad\qquad+41793851571836 q^5 +5687776756674922 q^6\\
&\qquad\qquad+415537010749335384 q^7+19979493668449934710 q^8+\cdots)
\end{split}
\ee
\be 
\begin{split}
n&=6:\\
\chi_0 &= q^{-\frac{19}{6}}(11-5092 q+1410294 q^2-329893048 q^3+80413387013 q^4\\
&\qquad\qquad-26136369675316 q^5+21403663588413994 q^6\\
&\qquad\qquad+15254385056035958248 q^7 +3241331580669093203257 q^8+\cdots)
\end{split}
\ee 
\be 
\begin{split}
n&=8:\\
\chi_0 &= q^{-\frac{25}{6}}(13-7540 q+2481630 q^2-635301160 q^3+147777747845 q^4\\
&\qquad\qquad -35198967511500 q^5+9781614373338990 q^6\\
&\qquad\qquad-3901008300552419160 q^7 +4120778511289519255280 q^8+\cdots)
\end{split}
\ee 

\subsubsection*{\underline{$n = 1$ {\rm mod} $2$ (type II)}}

For $n=1$ we have the first type II quasi-character in the $D_4$ series, with a single positive sign at the vacuum:
\be 
\begin{split}
n&=1:\\
\chi_0 &= q^{-\frac{2}{3}}(1 - 272 q - 34696 q^2 - 1058368 q^3 - 17332196 q^4 - 197239456 q^5 \\
&\qquad\qquad - 1749548096 q^6 - 12908725632 q^7 - 82505654138 q^8-\cdots)\\
\chi_1 &= q^{\frac{5}{6}}(1+52 q+1106 q^2+14808 q^3+147239 q^4+1183780 q^5\\
&\qquad\qquad +8095998 q^6+48688888 q^7+263508351 q^8+\cdots)
\end{split}
\ee 

\be 
\begin{split}
n&=3:\\
\chi_0 &= -q^{-\frac{5}{3}}(-1 + 296 q - 70412 q^2 + 33070784 q^3 + 11808560322 q^4\\
&\qquad\qquad +  1168760807520 q^5 + 61319216414024 q^6 + 2114915134928384 q^7\\
&\qquad\qquad +  53664123564854403 q^8+\cdots)
\end{split}
\ee 
\be 
\begin{split}
n&=5:\\
\chi_0 &= q^{-\frac{8}{3}}(3 - 1216 q + 309152 q^2 - 72227072 q^3 + 20981499344 q^4\\
&\qquad\qquad - 14708653418112 q^5 - 8731984857406080 q^6\\
&\qquad\qquad -  1519633405899254272 q^7 - 143608090423354740376 q^8-\cdots)
\end{split}
\ee 
\be 
\begin{split}
n&=7:\\
\chi_0 &= q^{-\frac{11}{3}}(13 - 6776 q + 2048612 q^2 - 497748224 q^3 + 115628448138 q^4 \\
&\qquad\qquad - 29998031865984 q^5 + 10841096859722744 q^6\\
&\qquad\qquad -  10162853918989450240 q^7 - 8455290382256521734421 q^8- \cdots)
\end{split}
\ee 

\subsection{$D_4$ class, $\ell=2$}

Here $k = 6n-2$, $n \in \mathbb{N}$

\subsubsection*{\underline{$n = 0$ {\rm mod} $2$ (type I)}}

For the value $n=2$, $c = 20$, we get the coset dual of the $D_{4,1}$ theory \cite{Gaberdiel:2016zke}. For the next value $n=4$, there characters are as follows:
\be 
\begin{split}
n&=4:\\
\chi_0 &= q^{-\frac{11}{6}}(5-572 q-72710 q^2+131993400 q^3+83188045875 q^4\\
&\qquad\qquad+12020096850740 q^5 +845062768380534 q^6\\
&\qquad\qquad+37194188193978200 q^7+1166609230664059875 q^8+\cdots)\\
\chi_1 &= q^{\frac{5}{3}}(33+8360 q+805300 q^2+43859200 q^3+1609921170 q^4+43986459264 q^5\\
&\qquad\qquad +952925556440 q^6+17093841177600 q^7+261962497543175 q^8+\cdots)
\end{split}
\ee 

\be 
\begin{split}
n&=6:\\
\chi_0 &= q^{-\frac{17}{6}}(3-748 q+74290 q^2+9327016 q^3-10453584189 q^4\\
&\qquad\qquad+12828238152012 q^5+10966314857286386 q^6\\
&\qquad\qquad+2513243971214727960 q^7+297247121423718739700 q^8+\cdots)
\end{split}
\ee
\be 
\begin{split}
n&=8:\\
\chi_0 &= q^{-\frac{23}{6}}(143-53636 q+10028230 q^2-955277400 q^3-119469184395 q^4\\
&\qquad\qquad+119516080787708 q^5-75409367064392350 q^6\\
&\qquad\qquad+101326663640492021160 q^7+110324231485901166793675 q^8+\cdots)
\end{split}
\ee
\be 
\begin{split}
n&=10:\\
\chi_0 &= q^{-\frac{29}{6}}(17-8468 q+2180974 q^2-366652104 q^3+34305961365 q^4\\
&\qquad\qquad +4283099167460 q^5-4092115890878682 q^6+2126167264602031592 q^7\\
&\qquad\qquad -1328680727438606492958 q^8+\cdots)
\end{split}
\ee

\subsubsection*{\underline{$n = 1$ {\rm mod} $2$ (type II)}}

For $n=3$, $c = 32$, we have type II behaviour of the quasi-characters as follows
\be 
\begin{split} 
n&=3:\\
\chi_0 &= q^{-\frac{4}{3}}(1 - 32 q - 65040 q^2 - 34608000 q^3 - 3498536840 q^4\\
&\qquad\qquad -  165008330944 q^5  - 4853582772800 q^6 - 102602094503680 q^7\\
&\qquad\qquad -  1692346088880100 q^8-\cdots)\\
\chi_1 &= q^{\frac{7}{6}}(1+188 q+12310 q^2+447800 q^3+11022585 q^4+204181636 q^5\\
&\qquad\qquad +3038074790 q^6+37922709320 q^7+409497231200 q^8+\cdots)
\end{split}
\ee 

For higher values of $n$, one finds odd numbers of positive signs that increase with $n$:
\be 
\begin{split}
n&=5:\\
\chi_0 &= q^{-\frac{7}{3}}(1 - 184 q + 5636 q^2 + 8297344 q^3 - 10602191110 q^4 - \\
&\qquad\qquad 7850068006464 q^5 - 1468949083484200 q^6 - 138283893885379072 q^7 - \\
&\qquad\qquad 8234555571519717877 q^8-\cdots)
\end{split}
\ee
\be 
\begin{split}
n&=7:\\
\chi_0 &= q^{-\frac{10}{3}}(11 - 3440 q + 505800 q^2 - 15332800 q^3 - 21223732860 q^4 \\
&\qquad\qquad + 15046714114208 q^5 - 19090416939480640 q^6\\
&\qquad\qquad -  18542226132988176000 q^7 - 5021426414672893297750 q^8-\cdots)
\end{split}
\ee
\be 
\begin{split}
n&=9:\\
\chi_0 &= q^{-\frac{13}{3}}(5 - 2184 q + 487580 q^2 - 66462400 q^3 + 2006696550 q^4 \\
&\qquad\qquad +  2715768684960 q^5 - 1644490529243752 q^6 + 1013680299336947200 q^7  \\
&\qquad\qquad - 1454063638609646721175 q^8-\cdots)
\end{split}
\ee

\bibliographystyle{JHEP}

\bibliography{towards}

\end{document}